\documentclass[12pt,preprint]{aastex}
\usepackage{aastexug}

\newcommand{\nch}{$N_\mathrm{ch}$}
\newcommand{\nmu}{$N_{\mu}$}
\shorttitle{Search for High Energy Neutrino Sources}
\shortauthors{J. Ahrens et al.}
\begin{document}
\title{Search for Point Sources of High Energy Neutrinos with AMANDA}
\author{
J.~Ahrens\altaffilmark{12},
X.~Bai\altaffilmark{1}, 
G.~Barouch \altaffilmark{8}, 
S.~W.~Barwick\altaffilmark{6}, 
R.~C.~Bay\altaffilmark{5},
T.~Becka\altaffilmark{12},  
K.-H.~Becker\altaffilmark{13}, 
D.~Bertrand\altaffilmark{10},
F.~Binon\altaffilmark{10}, 
A.~Biron\altaffilmark{2}
% ...
\newline
(The AMANDA Collaboration)}
\begin{abstract}
This paper describes the search for astronomical sources of high-energy neutrinos
using the AMANDA-B10 detector, an array of 302 photomultiplier tubes, used for the detection
of Cherenkov light from upward traveling neutrino-induced muons,
buried deep in ice at the South Pole.
The absolute pointing accuracy and angular resolution were studied by using coincident 
events between the AMANDA detector and two independent telescopes on the surface,
the GASP air Cherenkov telescope and the SPASE extensive air shower array.
Using data collected from April to October of 1997 (130.1 days of livetime), a general
survey of the northern hemisphere revealed no statistically significant excess of events
from any direction.
The sensitivity for a flux of muon neutrinos is based on the effective detection area for through-going muons.  Averaged over the Northern sky, the effective detection area exceeds 10,000~m$^2$ for $E_{\mu}\approx10$~TeV.
Neutrinos generated in the atmosphere by cosmic ray interactions were used to verify the predicted performance of the detector.
For a source with a differential energy spectrum proportional to $E_{\nu}^{-2}$ and declination larger than +40$^{\circ}$, we obtain 
$E^2(dN_{\nu}/dE)\leq 10^{-6}$~GeVcm$^{-2}$s$^{-1}$ for an energy threshold of 10 GeV.

\end{abstract}
\altaffiltext{1}{Bartol Research Institute, University of Delaware, Newark, DE 19716, USA}
\altaffiltext{2}{DESY-Zeuthen,D-15735, Zeuthen, Germany}
\altaffiltext{3}{Dept. of Technology, University of Kalmar, S-39182, Kalmar, Sweden}
\altaffiltext{4}{Fysikum, Stockholm University, S-11385, Stockholm, Sweden}
\altaffiltext{5}{Dept. of Physics, University of California, Berkeley, CA 94720 USA}
\altaffiltext{6}{Dept. of Physics and Astronomy, University of California, Irvine, CA 92697 USA}
\altaffiltext{7}{Dept. of Physics, Pennsylvania State University, University Park, PA 16802 USA}
\altaffiltext{8}{Dept. of Physics, University of Wisconsin, Madison,  WI 53706 USA}
\altaffiltext{9}{Lawrence Berkeley National Laboratory, Berkeley, CA 94720 USA}
\altaffiltext{10}{Universite Libre de Bruxelles, Science Faculty CP230, Boulevard du Triomphe, B-1050, Brussels, Belgium}
\altaffiltext{11}{Division of High Energy Physics, Uppsala University, S-75121, Uppsala, Sweden}
\altaffiltext{12}{Institute of Physics, University of Mainz, Staudinger Weg 7, D-55099, Mainz, Germany}
\altaffiltext{13}{Fachbereich 8 Physik, BUGH Wuppertal, D-42097 Wuppertal Germany}
\altaffiltext{14}{Physics Dept., University of Wisconsin, River Falls,  WI 54022 USA}
\altaffiltext{15}{Vrije Universiteit Brussel, Dienst ELEM, B-1050, Brussels, Belgium}
\altaffiltext{16}{University of Mons-Hainaut, Mons, Belgium}
\altaffiltext{17}{Dept. F\'{\i}sica, University Sim\'on Bol\'{\i}var, Caracas, Venezuela}

\altaffiltext{18}{Present address: CERN, CH-1211, Geneve 23, Switzerland}
\altaffiltext{19}{Present address:  Johns Hopkins University, Applied Physics Laboratory, Laurel, MD 20723, USA}
\altaffiltext{20}{Present address: Optical Networks Research, JDS Uniphase, 100 Willowbrook Rd., Freehold, NJ 07728-2879, USA}
\section{INTRODUCTION}
\label{intro}
\setcounter{footnote}{0}
Nature provides precious few information carriers from the deep recesses of space, and it is imperative to develop techniques to exploit each one. Throughout history, the photon messenger has made vital contributions to the understanding of the observable Universe. In this paper, we present results from a new generation of telescope designed to detect a very different kind of information carrier, high energy neutrinos (where $E_{\nu}>$ 1 TeV).  The search for astronomical sources of high energy neutrinos 
is one of the central missions of the Antarctic Muon and Neutrino Detector Array (AMANDA) \citep{Aman99}.
In this paper, we describe a general search for continuous\footnote{Although the
flux limits reported in this paper are computed assuming continuous emission,
upper bounds could be generated for periodic or episodic emission as well.}
emission from a spatially localized direction in the northern sky, restricted to
declinations greater than +5$^{\circ}$.
The search technique is conceptually simple:
a point source would be identified by a statistically significant enhancement over
expected background fluctuations from a particular direction.  Expected background is readily obtained experimentally from off-source sky bins within the same band of declination.
In contrast, unresolved, or diffuse signals, are characterized by an isotropic distribution and backgrounds are estimated by detector simulation programs.
The most favorable flux predictions for point sources are several orders of magnitude
lower than the most optimistic predictions for diffusely distributed sources.
However, atmospheric neutrino background is diffusely distributed as well, so the level of
intrinsic background in the diffuse search is also several orders of magnitude higher.
While signal-to-noise considerations favor the search for diffuse emission over point source searches,
the interpretation of a diffusely distributed signal is more ambiguous.
Thus, the search for point sources complements the search for diffuse sources.
The latter search is described in \citet{Hill}.  
The more specific searches for point emission from Gamma Ray Bursters \citep{AmanGRB}
and quasi-pointlike emission from galactic dark matter trapped in the core of the earth
\citep{AmanWIMP} are presented in separate papers since those
analyses were optimized for different flux spectra and different background characteristics.

\section{MOTIVATION}
\label{motivation}

The origin of cosmic rays is one of the oldest puzzles in particle astrophysics.
Shocks from galactic supernovae are widely believed to accelerate cosmic rays to
$\sim$10$^{15}$~eV, while the sources of cosmic rays at the most extreme energies are not known. 
Plausible models of particle acceleration exist for many classes of galactic and extra-galactic objects,
but supporting evidence for any model is largely circumstantial.
The observation of high-energy neutrinos from point sources would unequivocally confirm the
hadronic nature of such accelerators.
Unfortunately, the predicted neutrino fluxes from galactic and extra-galactic point sources
are too low to be detected with AMANDA-B10, although uncertainties in the model parameters
lead to considerable variation in the flux predictions. 

Supernova remnants (SNR) are one of the few classes of galactic sites that have the
capability to supply sufficient power to accelerate the galactic cosmic rays.
The diffusive shock mechanism naturally produces a power law spectrum of
$dN/dE$ $\propto E^{-2.1}$, which is consistent with the deduced spectral
index of cosmic rays. 

Recent observations of TeV gamma rays from plerions such as the Crab Nebula and SNR provide 
direct evidence for particle acceleration to high energies.
However, these observations do not provide compelling evidence for \textit{hadronic} acceleration 
due to an unfortunate ambiguity: it is possible (and even probable) that electrons are solely 
responsible for the high energy gamma-ray production.
But if SNR are the accelerators of galactic cosmic rays, they must also accelerate hadrons.
A class of models exploits this idea by suggesting that both protons and electrons are accelerated
by the supernova shock. 
Pions, both neutral and charged, are produced in the nuclear collisions between protons and
ambient material (a cosmic equivalent of a ``beam dump'' commonly used by terrestrial accelerators), 
and then decay to high-energy gamma rays and neutrinos. 

While the notion of particle acceleration by supernova shocks provides a credible and largely
consistent picture, not all observations neatly fit this scheme.
Alternative sites for cosmic ray acceleration may emerge from a detailed study of the neutrino sky.
For example, galactic microquasars --
a subclass of X-ray binary systems that exhibit relativistic radio jets --
have been identified as possible sources of high-energy neutrino emission 
\citep{Levinson01} and potential sources of the highest energy cosmic rays.
If they accelerate cosmic rays to high energies, then their dense environment creates suitable
conditions for an efficient beam dump.

Turning to extra-galactic sources, active galactic nuclei (AGNs) are among the most luminous objects
in the Universe and promising sources of neutrinos.
In these models, high-energy neutrino fluxes are generated near the central engine or in the jets
of radio-loud AGNs (\textit{e.g.}, blazars, a class of objects where the jet intersects the line of sight
of the observer).
The fact that gamma-ray emission has been detected \citep{Cantanese99} from nearby blazars Markarian (Mkn)
421 and 501\footnote{The review by \citet{Cantanese99} presents a current list of detected 
VHE gamma-ray sources.}  provides strong evidence for particle acceleration to high energies.
The time-averaged energy spectrum from Mkn 501 during 1997 is consistent with an unbroken power-law
energy spectrum up to 10~TeV \citep{TeVgamma,Konopelko99}. 
Beam dump models of neutrino production predict comparable fluxes of gamma-ray photons and neutrinos. However, gamma-ray photons at TeV energies may interact with material or photon fields in the
source, or interact with the diffuse infrared background photons during their flight.
Due to this reprocessing, the measured energy spectrum for gamma-ray photons may not trace the
energy spectrum of the source.
Consequently, it is possible for the ratio of neutrino flux to gamma-ray flux from a given source to
exceed unity.
Constraints on this ratio are discussed in Section~\ref{limits}.

Recently, it has been argued \citep{Buckley99} that the rapid time variability of high-energy 
photon emission from AGN blazars and the correlated variation between the X-ray and TeV regimes
disfavor hadronic acceleration models for this particular class of objects, but others have shown
that rapid and correlated variability can be accommodated by modest extensions to the existing
hadronic acceleration models \citep{rachen00,Dermer00}.
The vigorous debate suggests that high-energy neutrino detectors can play a central role in 
deciphering the acceleration mechanism. 
 
\begin{figure}
\plotone{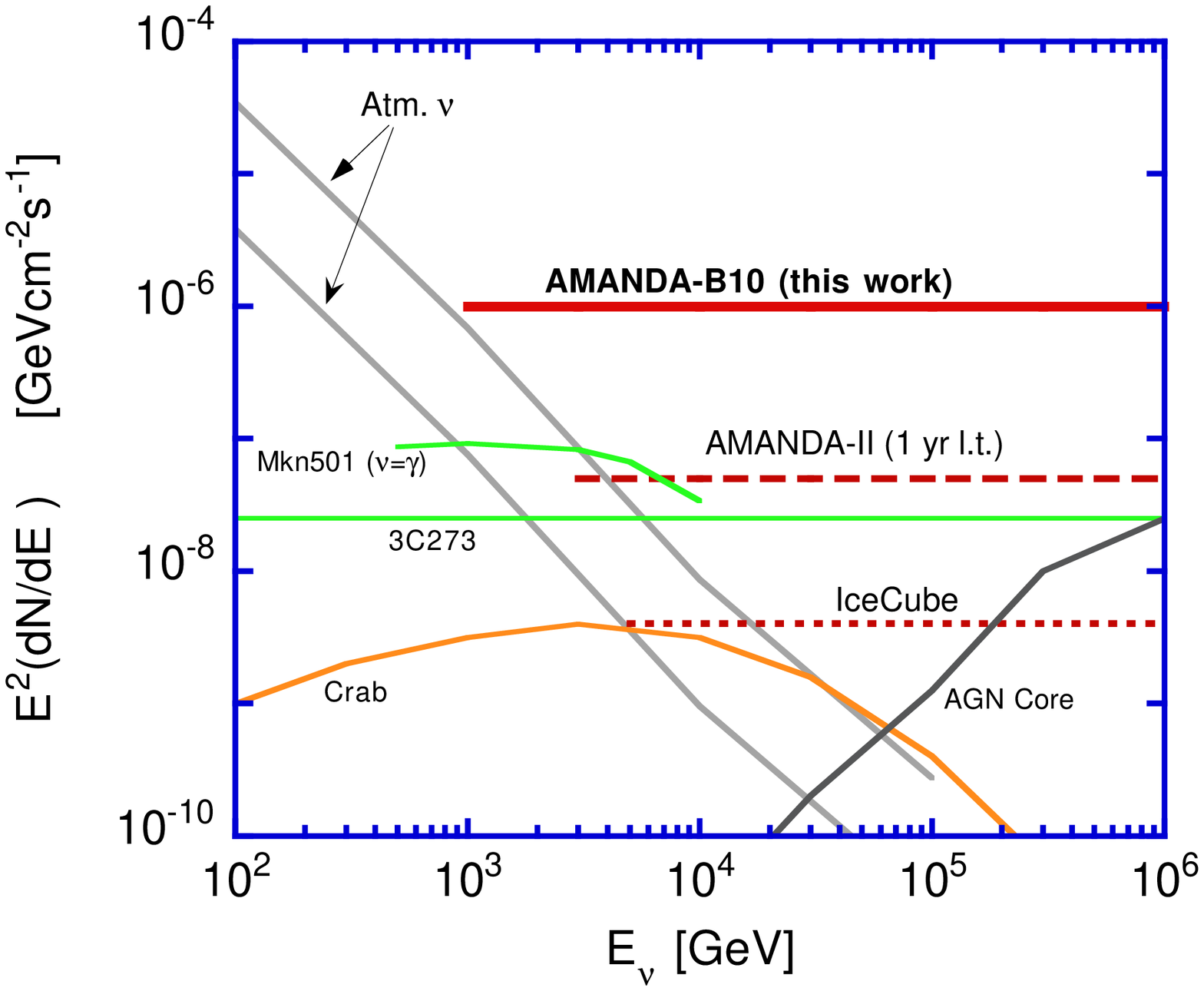}
\caption{Representative survey of $\nu +\overline{\nu}$ flux predictions from cosmic accelerators
of high-energy neutrinos. The AMANDA-B10 result is presented here. The dashed horizontal curves
give preliminary estimates of the minimum detectable flux by AMANDA-II after one year of live-time (1 yr l.t.)\citep{Barwick01} and 
IceCube \citep{Spiering01}. The atmospheric neutrino fluxes \citep{atm_nu_pred} are appropriate for
a circular patch of $1^{\circ}$ (lower) and $3^{\circ}$ radius. The curves do not include the normalization
uncertainty, possibly 30\% in magnitude \citep{Gaisser_atm}. Models: 3C273 \citep{Nellen}, 
Crab-Model I \citep{Bednarek97}, AGN core \citep{SS96}, and Mkn 501 assuming neutrino spectrum is identical
to observed gamma spectrum during flaring phase \citep{TeVgamma}.}
\label{fig:pointlim}
\end{figure}

Figure~\ref{fig:pointlim} provides a survey of model predictions for fluxes of high-energy neutrinos.
The models were selected to highlight the variation in spectral characteristics.
The curve labeled 3C273 is representative of several recent neutrino flux predictions,
\textit{e.g.}, a similar differential flux is predicted for microquasars \citep{microQ02} in the 
region of sky visible to AMANDA and the flat-spectrum radio quasar 3C279 \citep{Dermer01}.
The figure also shows the differential neutrino flux limit for an assumed source spectrum 
proportional to $E^{-2}$ for AMANDA-B10, and the anticipated corresponding sensitivity of AMANDA-II and IceCube.
The AMANDA-B10 result, the subject of this paper, is valid for declinations greater than +40$^{\circ}$. 
Many theoretical models of potential astronomical neutrino sources predict a very hard
energy spectrum, approximately $E^{-2}$ \citep{LMrev}, which leads to a most probable
energy for a detected neutrino well above 1~TeV (typically 10-30~TeV).
This high energy is a consequence of three facts: the neutrino cross-section for weak interactions
increases with neutrino energy, and the propagation length of the secondary muon increases
and the effective detection area increases as light emission along the muon track increases with energy.

Even though the cosmic ray puzzle provides a powerful motivation to explore the sky for neutrino
emission, not all high-energy neutrino sources need to contribute to the cosmic ray flux.
In particular, a powerful accelerator may be surrounded by too much material to emit high energy
photons or cosmic rays (they would interact and cascade down to lower energies), but this accelerator
could be discovered by exploiting the neutrino messenger.
Several models of such ``hidden'' sources have appeared in the literature.
For example, the predicted flux of neutrinos from pre-AGN objects \citep{hidden} leads to a muon 
detection rate of $\sim$10~y$^{-1}$km$^{-2}$.

\section{DESCRIPTION OF THE AMANDA DETECTOR}
\label{description}

\begin{figure}
\plotone{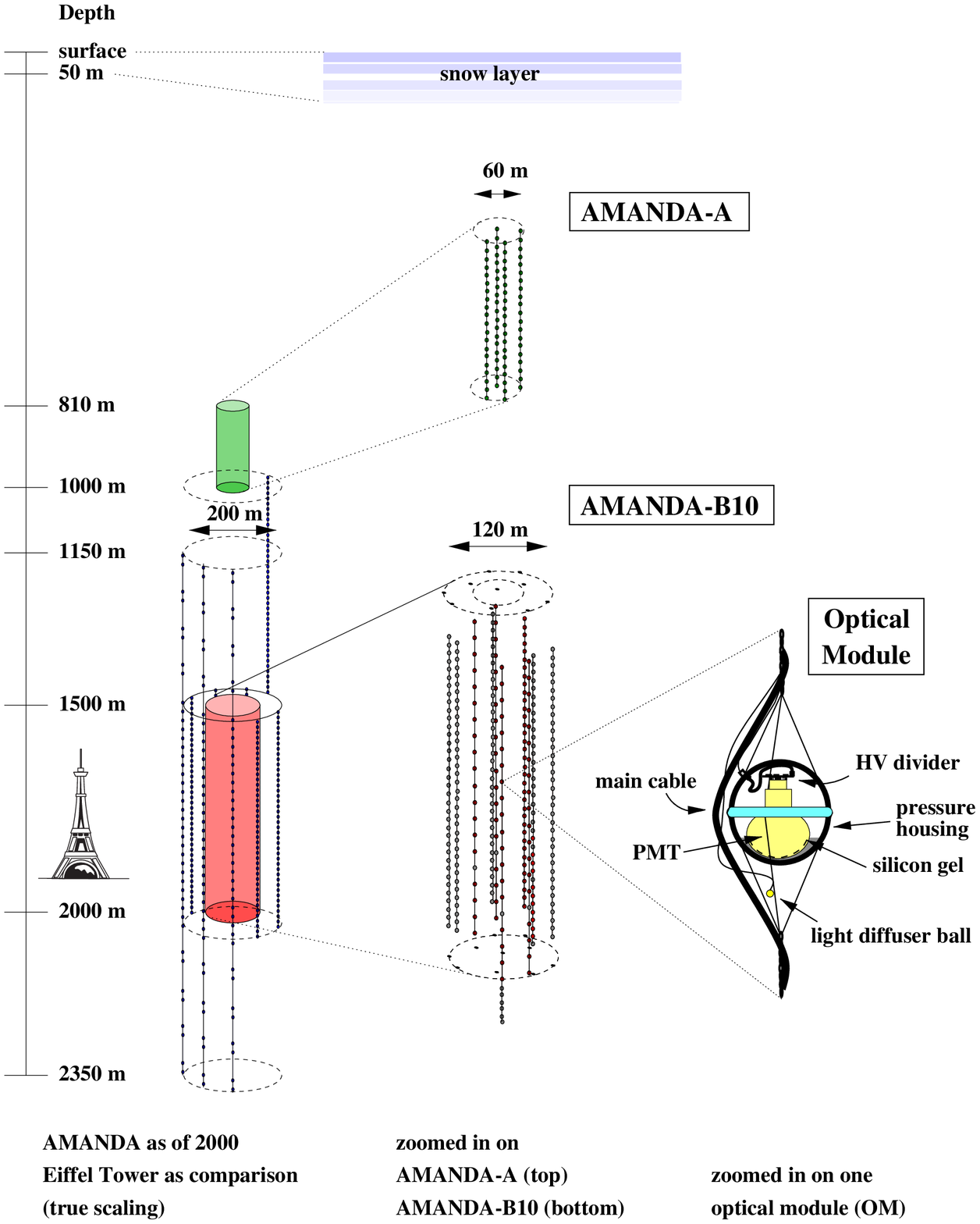}
\caption{Schematic view of the AMANDA neutrino telescope.
This paper describes an analysis of data taken in 1997 with AMANDA-B10,
the ten inner strings shown in the expanded view in the center.
Each dot represents one optical module in the array.}
\label{fig:AmII}
\end{figure}

The AMANDA telescope is located below the surface of the Antarctic ice sheet at the geographic south pole.
The neutrino detection technique relies on the detection of Cherenkov light from upward traveling 
neutrino-induced muons.
Figure~\ref{fig:AmII} shows the current configuration of the AMANDA detector.
The shallow array, AMANDA-A, was deployed to depths between 800 to 1000 m in
an exploratory phase of the project.
The deeper array of ten strings, referred to as AMANDA-B10, was deployed during
the austral summers between 1995 and 1997, to depths between 1500 and 2000 m.
At this depth, the optical properties are suitable for track reconstruction \citep{ice}.
The strings are arranged in a circular pattern when viewed from the surface.
The instrumented volume of AMANDA-B10 forms a vertical cylinder with a diameter
of 120 m.
Most electronics are housed on the surface in a research facility located within
a kilometer of the Amundsen-Scott South Pole Station.
The detector was commissioned in February 1997 \citep{Aman99,TAUP99} and expanded
by adding nine strings of OMs between December, 1997 and January, 2000.
The composite array of 19 strings forms the AMANDA-II array, which was 
commissioned in February of 2000.

AMANDA-B10 consists of 302 optical modules (OMs) arranged on ten vertical strings.
Each OM contains an 8-inch diameter
photomultiplier tube (PMT) controlled by passive electronics and housed in a glass
pressure vessel.
The OMs are connected to the surface by dedicated electrical cables,
which supply high voltage and carry the anode signals from the PMTs.
For each event, the amplitudes and arrival times of the pulses from the OMs
are digitized by peak ADCs and TDCs.
The TDCs are capable of measuring eight distinct pulses per channel.
The precision of the arrival time measurement is 5 ns.
Details of deployment, timing resolution, and detector operation can be found
in  \citet{Andres00,Nature01}.
Readout of the entire array was triggered by a majority logic system, which demanded that
at least 16 OMs produce signals, or ``hits'', within a time window of 2.2 $\mu$s.
This window takes into account the rather large time variation introduced by the
large geometric size of the detector and the cable propagation delays.
Random signals from the OMs (or noise) were observed at a rate of 300~Hz on the
inner four strings and 1.5~kHz for OMs on the outer six strings,
the difference being due to different levels of radioactive potassium in the glass
pressure vessels.
On average, random noise contributed one count per event to the majority trigger.

Optical absorption and scattering properties of the glacial ice that encapsulates
the AMANDA detector have been studied using light sources buried with the strings
and Cherenkov light from atmospheric muons.
These studies \citep{ice} confirm that the ice is not homogeneous, but consists of
horizontal strata correlated with climatological events in the past, such as ice 
ages \citep{ice2}.
Variations in the concentration of insoluble impurities between the strata
produce a strong modulation of the optical properties.
The absorption length, averaged over depth within the AMANDA-B10 array, is 110 m at a wavelength
of 400~nm, and the average effective scattering length is approximately 20 m. 

The detection of neutrinos relies on the observation of Cherenkov photons generated by
muons created in charged-current interactions.
At the energies of interest, muons typically propagate for distances in excess of several
kilometers (e.g. a muon with E = 10 TeV will travel 8 kilometers in water).
Therefore neutrino interactions outside the instrumented volume can be inferred
by the presence of a muon, providing a method to extend the volume of the detector
beyond the instrumented boundary of the array.
The average angle between the muon direction and the parent neutrino direction,
$\langle\theta_{\nu\mu}\rangle$, is approximately $0.65^{\circ}/(E_{\nu}/\mathrm{TeV})^{0.48}$ for $E_{\nu}$ less than 100 TeV \citep{munu01}.  However, nearly independent of the muon energy, the precision of the measured muon direction in AMANDA-B10 is approximately 4 degrees
(see Section~\ref{pointing}), which dominates the angular uncertainty in the neutrino direction.

The AMANDA-B10 data analysed here was collected between April and October of 1997.
Once construction was completed in February, 1997, calibration and data management
activities continued until April.
Operations ceased between late October, 1997 and February, 1998, due to
the beginning of construction of the AMANDA-II array.
Furthermore, limitations in the data acquisition and archiving system during
that first year of operation reduced the total livetime to approximately 130.1 days.

\section{SIMULATIONS}
\label{simulation}

Astronomical signals are unlikely to produce more than a few tens of upgoing neutrino events per
year in AMANDA-B10.
Data is therefore overwhelmingly dominated by two types of background:
downgoing atmospheric muons generate essentially all of the recorded events, and
atmospheric neutrinos contribute a few tens of events per day.
The point source search relies on a good understanding of both signal and background through simulations
based on Monte Carlo techniques.

Atmospheric muon events are generated from the measured flux of cosmic rays \citep{CRflux}.  
Two different air shower simulation packages were used to assess systematic uncertainty:
\texttt{BASIEV} \citep{Basiev}, and \texttt{CORSIKA} (version 5.6) \citep{CORSIKA} using the 
\texttt{QGSJET} hadronic interaction model.
\texttt{CORSIKA} was modified to include the curved geometry of the earth and atmosphere to provide a more
accurate description of the flux at zenith angles close to the horizon.
Most characteristics of the events generated with \texttt{BASIEV} were found to be similar to those
from the more accurate, but computationally more intensive, \texttt{CORSIKA} simulation.
The density profile of the atmosphere was modified for polar conditions, but no attempt was
made to replicate the small seasonal variations of the trigger rate \citep{Bouchta99}.
Muon tracking from the surface to the detector was handled by the muon propagation program
\texttt{MUDEDX} \citep{Lohmann} and the energy loss characteristics compared against two 
additional propagation programs which are available for general use: \texttt{PROPMU} \citep{LS} and 
\texttt{MMC} \citep{MMC}.
Integral lateral distributions of muons at the depth of AMANDA were simulated for proton and
iron showers \citep{SPAM02}, and used for verification of the detector performance as described
below.

The propagation of upward traveling muons from neutrino interactions was treated differently than
downgoing atmospheric muons because the energies of signal neutrinos were expected to extend to much
higher energy.
Neutrinos were tracked through the earth and allowed to interact in the ice within or near the instrumented
volume of the detector, or in the bedrock below \citep{Hill97}. 
Muons with energies above $10^{5.5}$ GeV at production were propagated using \texttt{PROPMU} until they
reached the rock-ice boundary and then propagated through the ice in exactly the same way as downgoing 
atmospheric muons.
For energies below $10^{5.5}$ GeV at the production vertex, muons were tracked with \texttt{MUDEDX}.
The background fluxes from atmospheric $\nu_{\mu}$ and $\overline{\nu}_{\mu}$ \citep{atm_nu_pred} were
included (AMANDA cannot differentiate the charge sign of the muon).  

In addition to background from atmospheric muons and muon neutrinos,
the detection efficiency for atmospheric electron neutrinos has been simulated.
At the relevant energies, the flux of $\nu_e$ is far smaller than for $\nu_{\mu}$,
so the background contribution is small \textit{a priori}.
Furthermore, the topology of electron neutrino events, reflecting the electromagnetic shower
generated by the secondary electron, is spherical rather than linear, and this characteristic
has been exploited to further increase the rejection.
Simulations show that the detection rate of atmospheric $\nu_e$ in the point source analysis is only 0.3\% of $\nu_{\mu}$
\citep{Young00} and therefore negligible in this context. We note that a separate analysis was devised to search for electron neutrinos and consequently achieved a much larger sensitivity\citep{cascade02}.

An overview of the simulation of the detector response is given in \citet{Hundertmark00}.
The depth dependence of the optical properties of the bulk ice \citep{ICRC99} is included
along with a realistic treatment of trigger conditions, intrinsic noise rates, and 
hardware-dependent pulse shapes.
The linearity and saturation of the photomultiplier response is included.
The angle-dependent sensitivity of the optical module was computed from a convolution
of PMT quantum and collection efficiencies as estimated by the manufacturer, 
detection efficiency, wavelength-dependent transparency through the pressure housing and buffer gel,
and obscuration by cables and mechanical support hardware.
The relative angular dependence of the sensitivity of the photomultiplier tubes was
measured in the laboratory.
The local optical properties of the refrozen ice were included in the photon tables
that describe the probability and the timing characteristics of photon propagation \citep{atm_nu02}.

\begin{figure}
\plotone{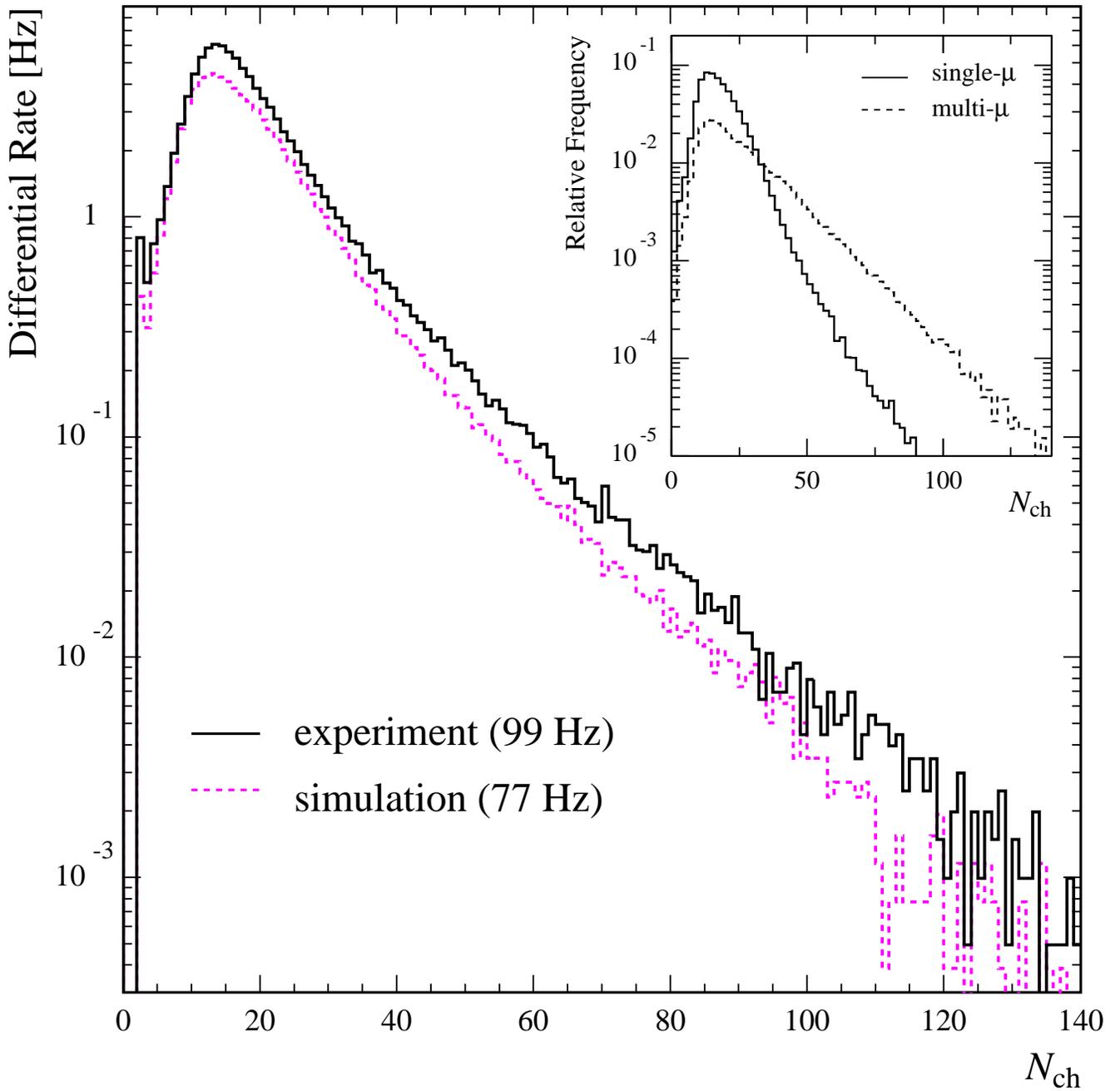}
\caption{Differential distribution of observed (solid line) and predicted (dashed line) trigger rates
as a function of the event multiplicity \nch\ (\textit{i.e.}, the number of optical modules that 
participate in each event). The integrated rates are given in parenthesis. Note that \nch\ extends below the majority logic threshold of 16 due to removal of data caused by experimental artifacts.
Inset: Relative contribution to the trigger rates from single muons (solid) and multiple-muon
bundles (dashed) that traverse the fiducial volume of the array.}
\label{fig:Nhit}
\end{figure}

Figure~\ref{fig:Nhit} presents the differential distribution of the multiplicity 
of optical modules, or channels, participating in an event, \nch, for the full detector
simulation and for experimental data after known detector-related artifacts were removed.
The integrated rates differ by less than 25\%, which is within the 
systematic uncertainties associated with the flux of the primary cosmic 
rays \citep{Gaisser_atm} and uncertainties associated with the absolute sensitivity of the optical module.
The agreement in shape demonstrates a good understanding of the overall 
sensitivity of the array for the most common events that trigger AMANDA-B10.

\begin{figure}
\plotone{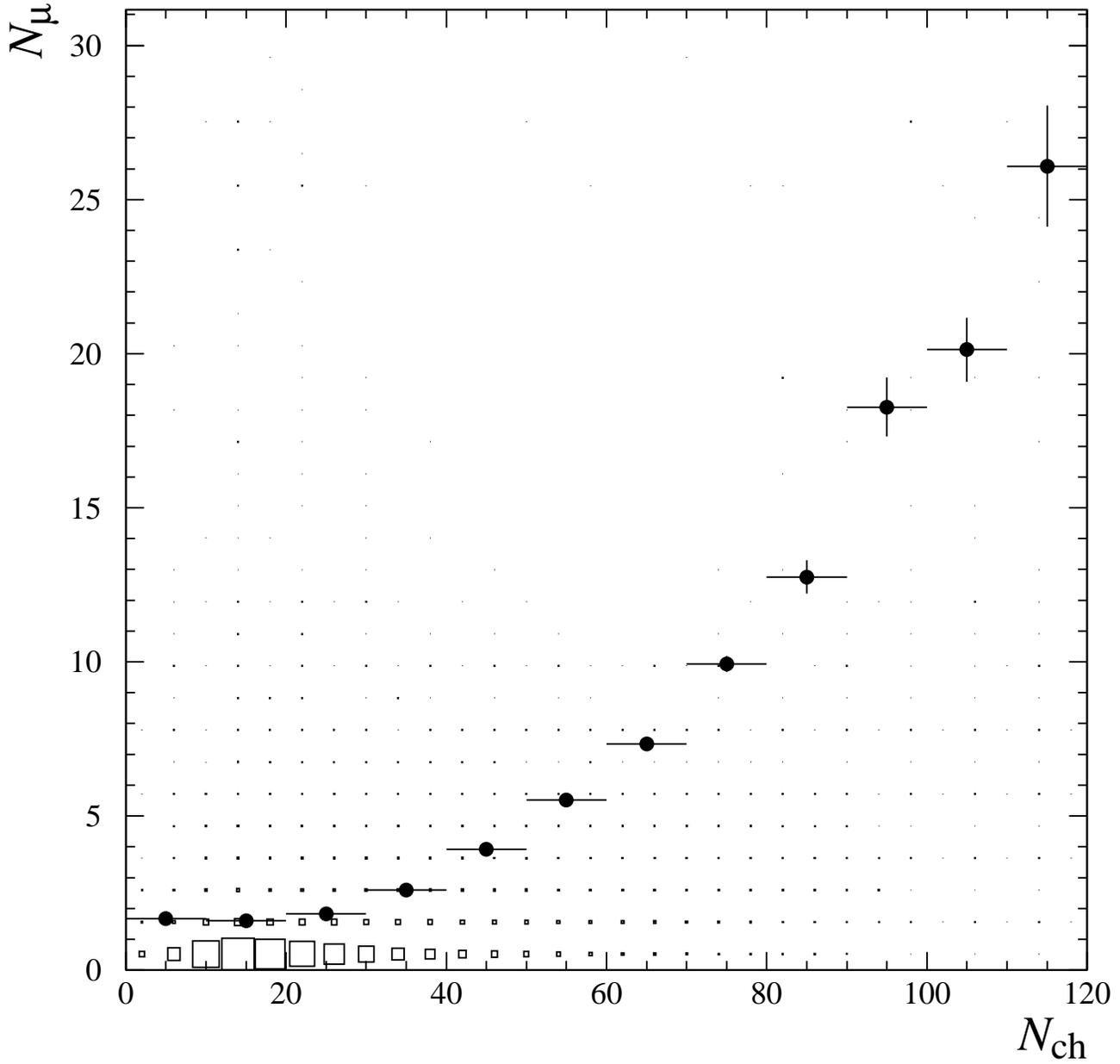}
\caption{Muon multiplicity, \nmu, versus OM multiplicity, \nch, from a full
detector simulation.
The average values (dots) show the correlation between these two quantities,
and the vertical error bars show the statistical uncertainty.}
\label{fig:Nmu}
\end{figure}

As the inset of Fig.~\ref{fig:Nhit} shows, the largest values of \nch\
are produced by events with more than one muon.
This information provides indirect evidence that the response of AMANDA to single
high-energy muons is correctly modeled, by the following argument.
Multi-muon bundles that reach AMANDA mainly consist of  muons below the critical energy of 600 GeV, which implies that energy loss due to ionization is near minimum. The Cherenkov light production from muons well above the critical energy is dominated by electromagnetic showers, and the total light from a muon with $E_{\mu}$ is approximately equal to the light of a bundle of N muons, where N$\sim E_{\mu}/E_\mathrm{crit}$.
Therefore, the light production by a multi-muon event can be related to the light
production by a single-muon event.
For example, the average energy loss per unit length for a muon with energy
$10^{13}$ eV is approximately a factor of 15 larger than for a single muon, as long as the energy is below 600 GeV.  
There are several modest limitations to this line of reasoning. 
One is that the lateral distribution of multi-muon events generates Cherenkov
photons over a much larger cylinder than a single muon.
Another difference is that multi-muon events deposit Cherenkov photons more
uniformly than the equivalent high energy muon, for which energy loss is dominated 
by occasional pair production and bremsstrahlung.
However, optical scattering by the ice mitigates the effects of non-uniform photon generation.
Simulations show that bundles of 20 muons generate a similar \nch\
distribution as single muons with an energy of 10~TeV.  
The correlation between muon multiplicity and \nch\ multiplicity is shown in Fig.~\ref{fig:Nmu}.

\section{ANALYSIS PROCEDURE}
\label{analysis}

The analysis procedure exploits two essential characteristics of the signal to
simplify the analysis relative to atmospheric neutrino measurements.
First, the sources are assumed to be point sources in the sky, so only events
within a restricted angular region are considered.
Secondly, we use the topological and directional characteristics of the spectrally
hard neutrino signal to help reject poorly reconstructed atmospheric muons 
(\textit{i.e.}, downward traveling muons reconstructed as upward traveling)
and atmospheric neutrinos, both of which have softer spectra.
Unlike many neutrino detectors, the effective sensitivity of AMANDA varies dramatically as
a function of the background rejection requirements.
By concentrating on harder spectra, the effective area of the detector can be
increased by relaxing the background rejection criteria.
Since the point source analysis tolerates a larger background ($B$) contamination in the final data sample,
the analysis procedure optimizes on signal to noise ($S/\sqrt{B}$) rather than signal purity ($S/B$).

Prior to track reconstruction and event selection, experimental data were selected from runs
that exhibited no abnormal behavior, and various instrumentally-induced artifacts were removed.  Once the data in the runs were certified, individual OMs in the array were examined to insure proper operation.
OM channels with hardware malfunction ($\sim$15\% of OMs), such as pickup from unusually
large external noise sources or fluctuations in the response of the amplifier electronics, were rejected. Approximately 85\% of OMs remained after deselection.
Occasional signals induced by cross-talk in the electrical cables or surface electronics exhibited characteristic
behavior and could be removed by straightforward restrictions on pulse amplitude and width.
Noise signals generated internally by the photomultiplier tubes were readily removed if their
time of arrival occurred earlier than 5 microseconds prior to the formation of the event trigger. The reconstruction programs stochastically account for PMT noise within the event duration. 

After this initial data cleaning, a number of event reconstruction techniques \citep{Andres00} 
are applied to the data. 
The most sophisticated technique relies on a search in multiparameter space to find
the maximum likelihood for a track hypothesis given the recorded hits.
After reconstruction is completed, events are selected according to a set of criteria that
retain only the highest-quality events that possess topological and directional information
consistent with those expected for upgoing neutrino-induced muons.
In a first step, the data sample of $1.05\times10^9$ events at trigger level is 
reduced to a more manageable size by two filtering stages.
Most events in data are readily identified as due to downward traveling muons by 
computationally fast reconstruction routines.
Removing these events reduces the data approximately by a factor of $10^3$.

An iterative analysis procedure was developed to maximize
$S/\sqrt{B}$
for a simulated signal with an energy spectrum proportional to $E^{-2}$. It ignored the absolute time of the event, which helped to minimize bias from potential sources in the data.  In this optimization the background was determined from experimental data
by assuming that the fraction of signal events in the data sample is negligible.
After the filtering stages, cuts were applied sequentially on a set of selection variables,
with several variables included more than once.
The specific value for each cut after stage 2 was chosen to retain $\gtrsim$~80\% of the signal.
At each stage, given this constraint on signal efficiency,
the same cut was made on data for the variable with the largest rejection power, 
$R = \epsilon_\mathrm{sig}/\epsilon_\mathrm{bgr}$,
where $\epsilon = N^\mathrm{pass}/N^{0}$,
and $N^{0}$ and $N^\mathrm{pass}$ are the number of events before and 
after the application of the selection cut respectively.
The signal-to-noise ratio was then computed as a function of zenith angle to ensure that
the acceptance of AMANDA-B10 remained as large as possible near the horizon.
The effective areas for detection of background and signal needed for this computation were
determined from simulations (as described in Section~\ref{limits}).
After each stage, this procedure was repeated on the remaining variables.

Besides restrictions on the reconstructed zenith angle, $\theta$,
the most effective selection criteria impose a threshold on the number, $N_\mathrm{dir}$,
of only slightly scattered, or ``direct'' photons (\textit{i.e.}, photons that travel between
the reconstructed track and the OM in nearly a straight line), 
and the track length, $L_\mathrm{dir}$, over which these photons are detected.
Furthermore, the analysis requires a minimum goodness-of-fit value from
the maximum likelihood procedure.
Other criteria evaluate the topological distribution of the photon emission
using variables that describe the granularity of the light pattern along the
trajectory, and a related observable that assesses the sphericity of the photon 
pattern\footnote{Muons normally generate a linear distribution of Cherenkov 
photons, whereas electromagnetic cascades initiated by pair production or
bremsstrahlung produce spherically symmetric distributions.}. 
Table \ref{tb:variables} shows the selection variables and cuts used in this
analysis, including a brief technical description of the two filtering stages.
The selection variables were introduced previously \citep{atm_nu02} and a
complete description is also available \citep{Young00}.
Also shown in the table are efficiencies and rejection factors at each stage
of the analysis for experimental data, simulated background and simulated
signal, averaged over all angles.

\begin{figure}
\plotone{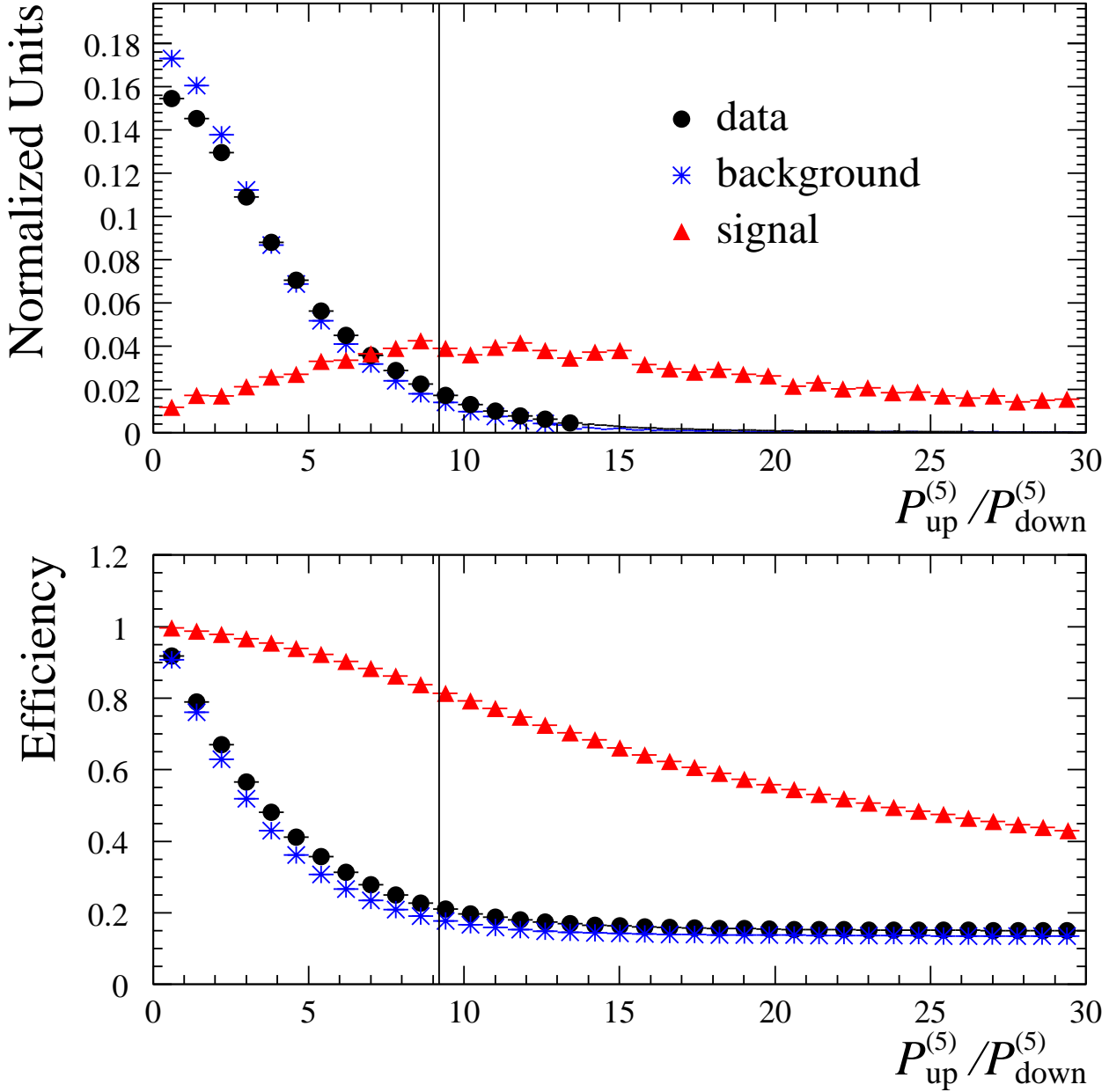}
\caption{Upper panel: equally normalized distributions for experimental data (circles) and
simulated background (asterisks) and signal (triangles) for stage 4 of the point source analysis,
which compares the best likelihood for an upgoing track hypothesis with the likelihood for a track
in the opposite (\textit{i.e.}, downgoing) direction.
Lower panel: passing efficiencies as a function of cut value. The efficiency is obtained from the integrated sums
of distributions shown in the upper panel from given value to infinity.  The vertical lines indicate the cut applied in the analysis.}
\label{fig:cut1}
\end{figure}

\begin{figure}
\plotone{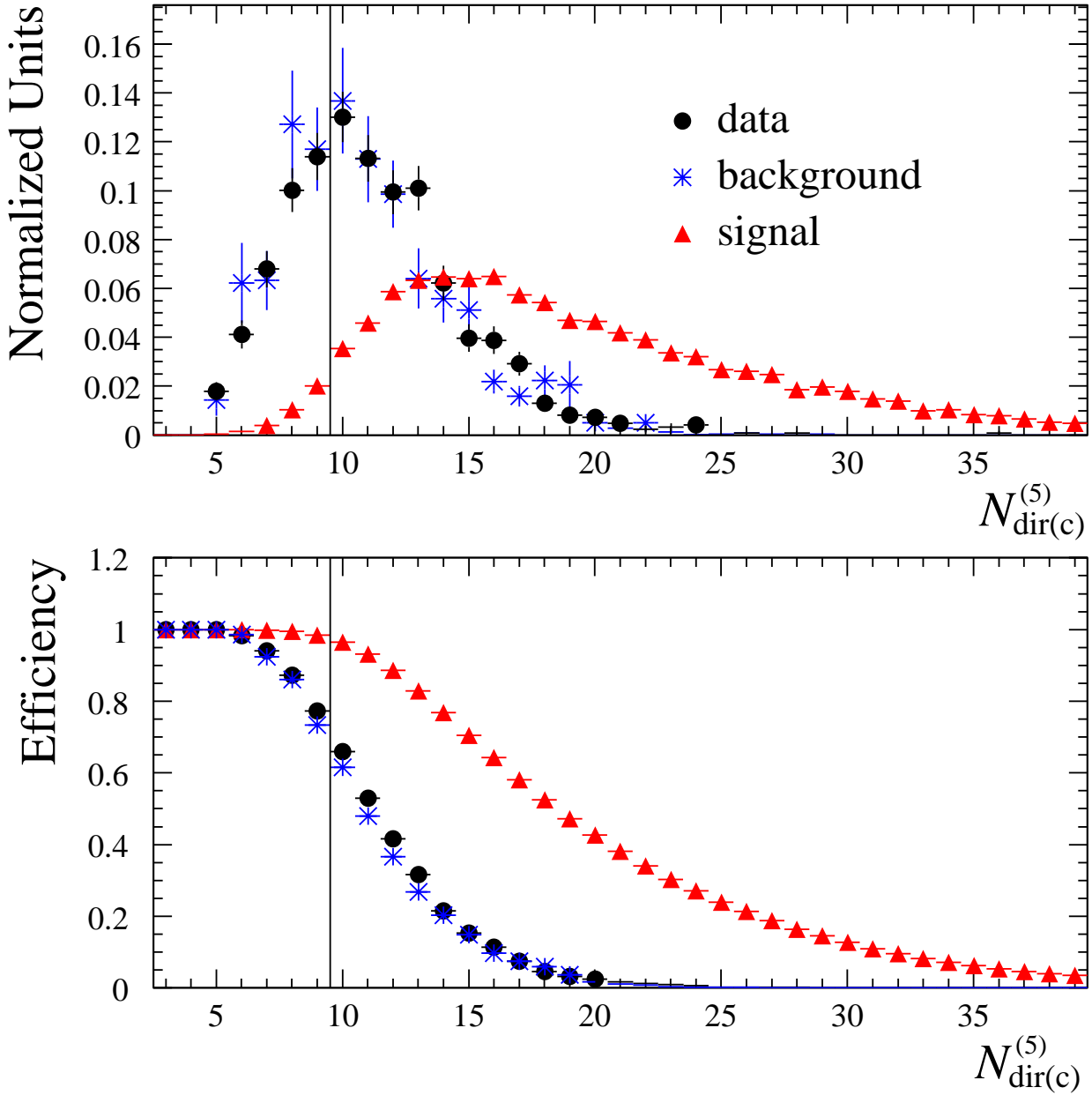}
\caption{Upper panel: similar to Fig.~\ref{fig:cut1}, but for analysis stage 13 which 
involves the number of direct hits in a time window of $-15$ to $+75$~ns.
Lower panel: Passing efficiencies -- defined as integrated sums, from given value to infinity,
of distributions shown in the upper panel -- as function of cut value.
The vertical lines indicate the cut applied in the analysis.}
\label{fig:cut2}
\end{figure}

The simulated background from atmospheric muons and neutrinos is compared to data
at all stages of the analysis to establish confidence in the simulation.
Figure~\ref{fig:cut1} shows the comparison at stage 4,
which is sufficiently early in the analysis to retain high statistical precision. 
Figure~\ref{fig:cut2} shows the final stage (13) of the analysis procedure. 
We note that the signal sensitivity determined from simulations is quite robust against small
deviations between the simulated event distribution and the actual response of the detector \citep{Young00}.

\begin{figure}
\plotone{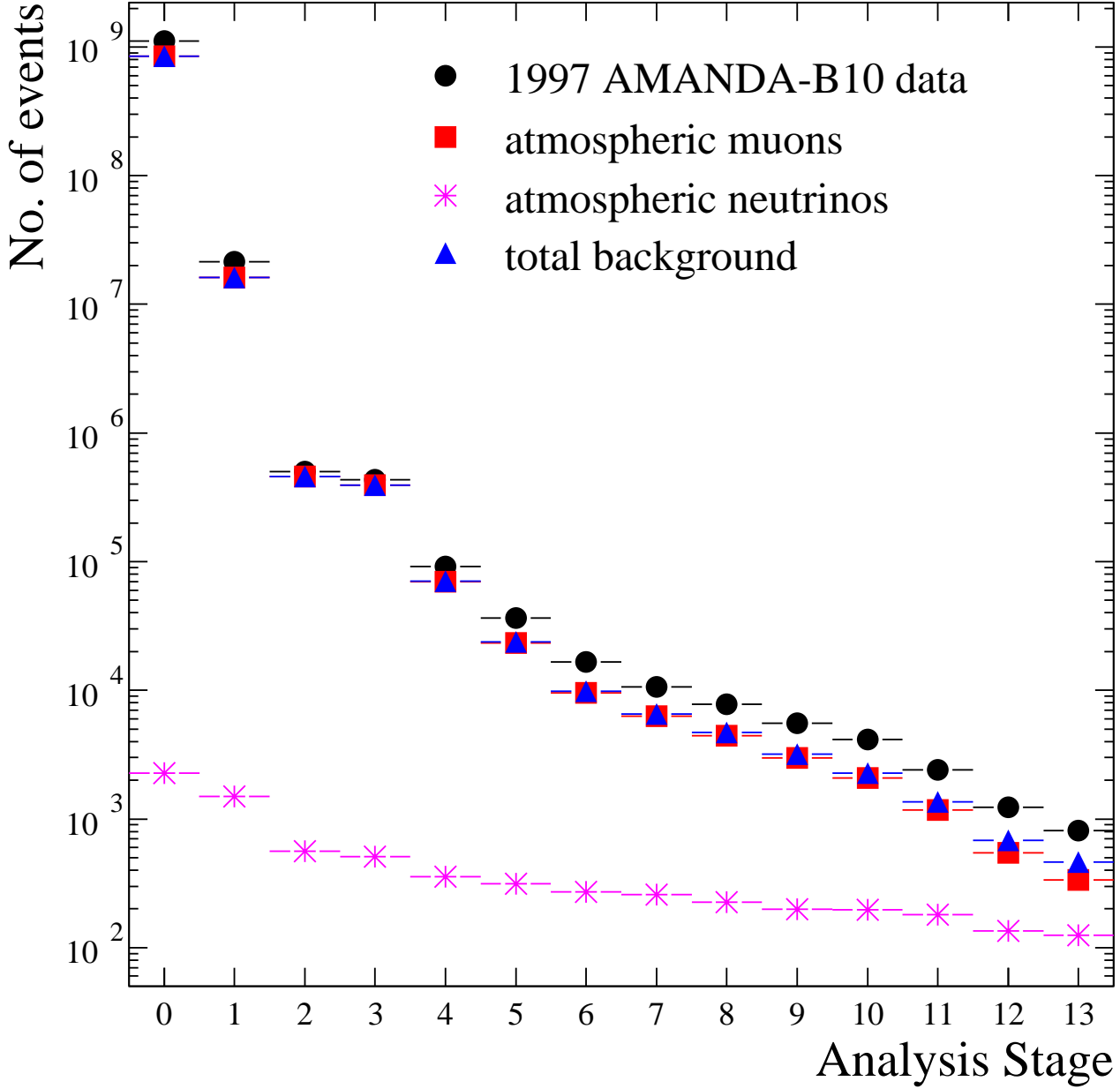}
\caption{Number of events remaining in the sample as the selection criteria in the 13 analysis stages 
listed in Table~\ref{tb:variables} are applied sequentially.
The 1997 AMANDA-B10 data (circles) are compared to simulated background from 
atmospheric muons generated by cosmic ray interactions (squares) and
from muons induced by atmospheric neutrinos (asterisks).
Also shown is the sum of atmospheric muon and neutrino backgrounds (triangles).
No normalization was applied, but systematic uncertainty at the trigger level (analysis stage 0) may be 
as large as $\pm$30\%.}
\label{fig:bg}
\end{figure}

Due to the large experimental data sample, precision studies of detector performance are possible
from the most common events in the sample to extremely rare components.
The predicted and experimental sample sizes are compared through all stages of the analysis
to establish the absolute calibration of detector sensitivity.
This method assumes that signal from astronomical sources contributes negligibly to the sample.
Figure~\ref{fig:bg} shows that the simulated background and data agree to within a factor two at all
stages of the analysis even though the size of the event samples vary by six orders of magnitude.
The relative rates and the agreements in shape of the selection variable distributions
provide evidence that the background generators and detector simulation programs are an adequate
description of the detector physics, including detector response. 
Due to the optimization on signal to noise, all stages of the analysis produce event
samples that are dominated by poorly reconstructed downgoing atmospheric muon events.
Diffuse backgrounds from atmospheric neutrinos become noticeable only after rejecting most of the
poorly reconstructed atmospheric muons, but they never dominate the event sample.

Obviously, atmospheric muon data is an imperfect tool to investigate the sensitivity of the
detector to neutrino-induced muons, due to the downgoing nature of the events and differences
in the energy and multiplicity distributions.
This concern is addressed by the measurement of atmospheric neutrinos \citep{Nature01,atm_nu02} 
which were used to verify the basic operational sensitivity of AMANDA-B10 to a known neutrino signal. 
In this analysis, the relative agreement between the measured and predicted event rates is 30\%, which is
consistent with uncertainties in the measured flux of cosmic-ray primaries, theoretical uncertainty 
in the interaction models, and systematic uncertainties in the modeling of the detector response. 
However, because of the steeply falling energy spectrum for atmospheric neutrinos,
the mean energy of the muons induced by charged-current interactions is close to the energy threshold
of the detector, which implies that they cannot be used to reliably probe the high-energy response of AMANDA-B10.

\section{POINTING RESOLUTION AND POINT SOURCE SEARCH}
\label{pointing}

\begin{figure}
\plotone{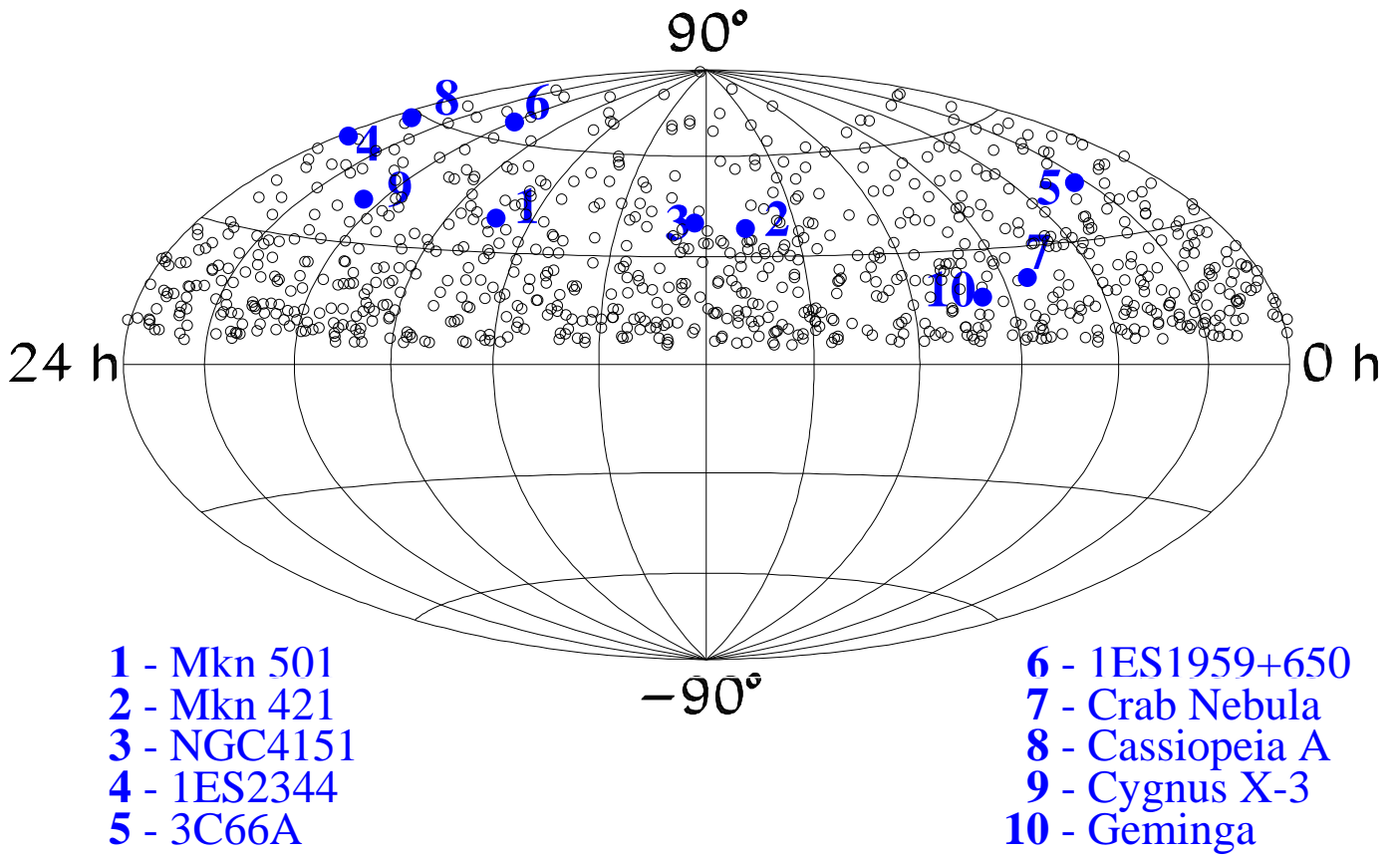}
\caption{Sky plot of 815 events obtained from the point source analysis.
Horizontal coordinates are right ascension and vertical coordinates are declination.
Also shown are the sky coordinates for ten potential high-energy neutrino sources.}
\label{fig:sky}
\end{figure}

\begin{figure} 
\plotone{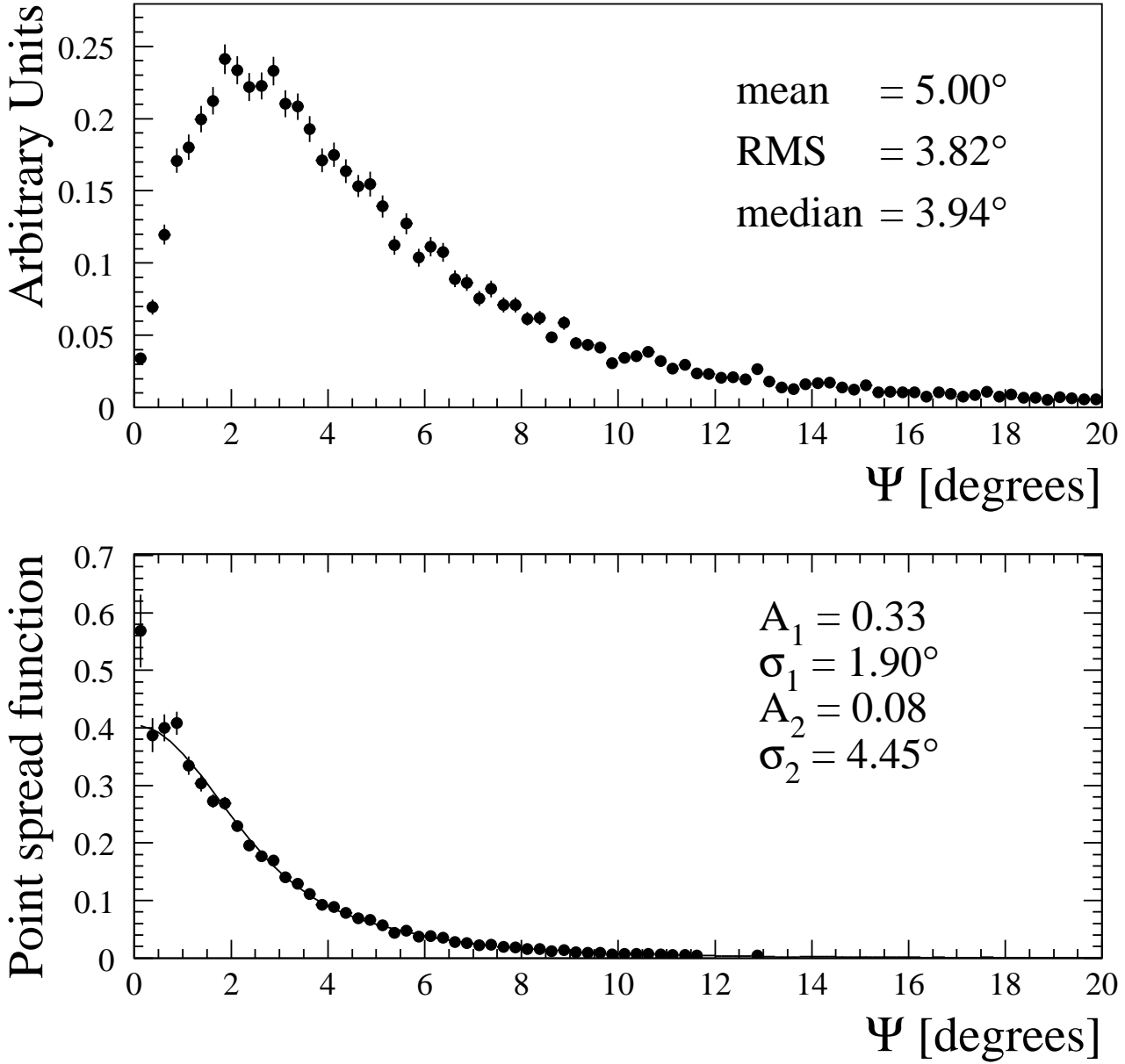}
\caption{Upper panel: Distribution of space angle, $\Psi$, between true and reconstructed muon 
track direction for simulated signal with neutrino energy spectrum proportional to $E^{-2}$, 
averaged over direction. 
Lower panel: Point spread function for signal in AMANDA-B10, deduced from the space angle distribution
in the upper panel. A function composed of the sum of two Gaussians was used to characterize this distribution. }
\label{fig:angleres}
\end{figure}

The final stage of the data analysis procedure yields a sample of 815 events (as is evident from Fig.~\ref{fig:bg}, atmospheric neutrinos contribute about 25\% of the events to the simulated background.).
Visual inspection of the distribution in the sky of the final event sample, shown in Figure~\ref{fig:sky},
reveals no obvious clustering.   
In order to perform a quantitative search for possible sources of high-energy neutrinos
in the northern hemisphere the sky was divided into non-overlapping angular bins of
approximately equal solid-angle coverage.
A point source would then be revealed by a statistically significant clustering of events within a
particular angular bin.
The optimal bin size and shape depend on the space angle resolution of the detector,
which can be expressed in terms of a point spread function.
The space angle deviation, $\Psi$,  between the true angular coordinates of a muon,
($\theta_\mathrm{true},\phi_\mathrm{true}$), and the reconstructed coordinates,
($\theta_\mathrm{rec},\phi_\mathrm{rec}$), is given by
\begin{equation}
\cos\Psi = \cos\theta_\mathrm{rec}\cos\theta_\mathrm{true} + 
\sin\theta_\mathrm{rec}\sin\theta_\mathrm{true}\cos(\phi_\mathrm{rec}-\phi_\mathrm{true}).
\end{equation}
Figure~\ref{fig:angleres} shows the distribution of $\Psi$ and the corresponding point spread function,
computed from the $\Psi$ distribution by dividing by the appropriate solid angle, for the simulated sample
of upward traveling muons generated by neutrinos with an $E^{-2}$ energy spectrum.
A median value of $\Psi = 3.9^{\circ}$, averaged over positive declinations, is achieved.
A function involving the sum of two Gaussian distributions was fit to the point spread function.  It yields an amplitude ratio of $A_2/A_1$ = 0.25,
indicating the importance of the second component related to the
degrading angular resolution at large muon energies \citep{Young00}.
Given this point spread function and the relatively small number of background events
shown in Fig.~\ref{fig:sky}, the optimal slice in zenith angle is 11.25 degrees \citep{Young00}.
For azimuth angle, a weak optimum occurs for a width of 12 degrees for the declination band closest to the horizon. 
These angular dimensions of the bins were chosen to maximize the signal to noise.

Two studies were performed to check the predicted space angle resolution and absolute
pointing accuracy.
The first uses AMANDA events that were also tagged by the GASP air Cherenkov
telescope \citep{GASP}.
GASP determines the direction of the air shower and AMANDA measures the direction of
the penetrating muon component.
At AMANDA depths, these events are almost entirely single muons. Unfortunately, the duty cycle of operation is low, so the sample size if relatively small.
To improve the statistical accuracy of the angular resolution studies, a second method based on extensive air showers
was developed.  This method utilized events that triggered both the SPASE
array and AMANDA \citep{SPAM02}.
SPASE responds to the electron and photon content of the shower front that reaches 
the surface.
Since the direction of muons within the air shower event is nearly perpendicular to the 
shower front, the difference between the direction of the air shower and 
the reconstructed muon direction can be used to deduce the angular resolution of AMANDA.    
SPASE measures the direction of an air shower with a pointing resolution of approximately
1-2 degrees \citep{SPASE} (depending on shower size), which is small enough to calibrate the AMANDA pointing resolution.

\begin{figure} 
\plotone{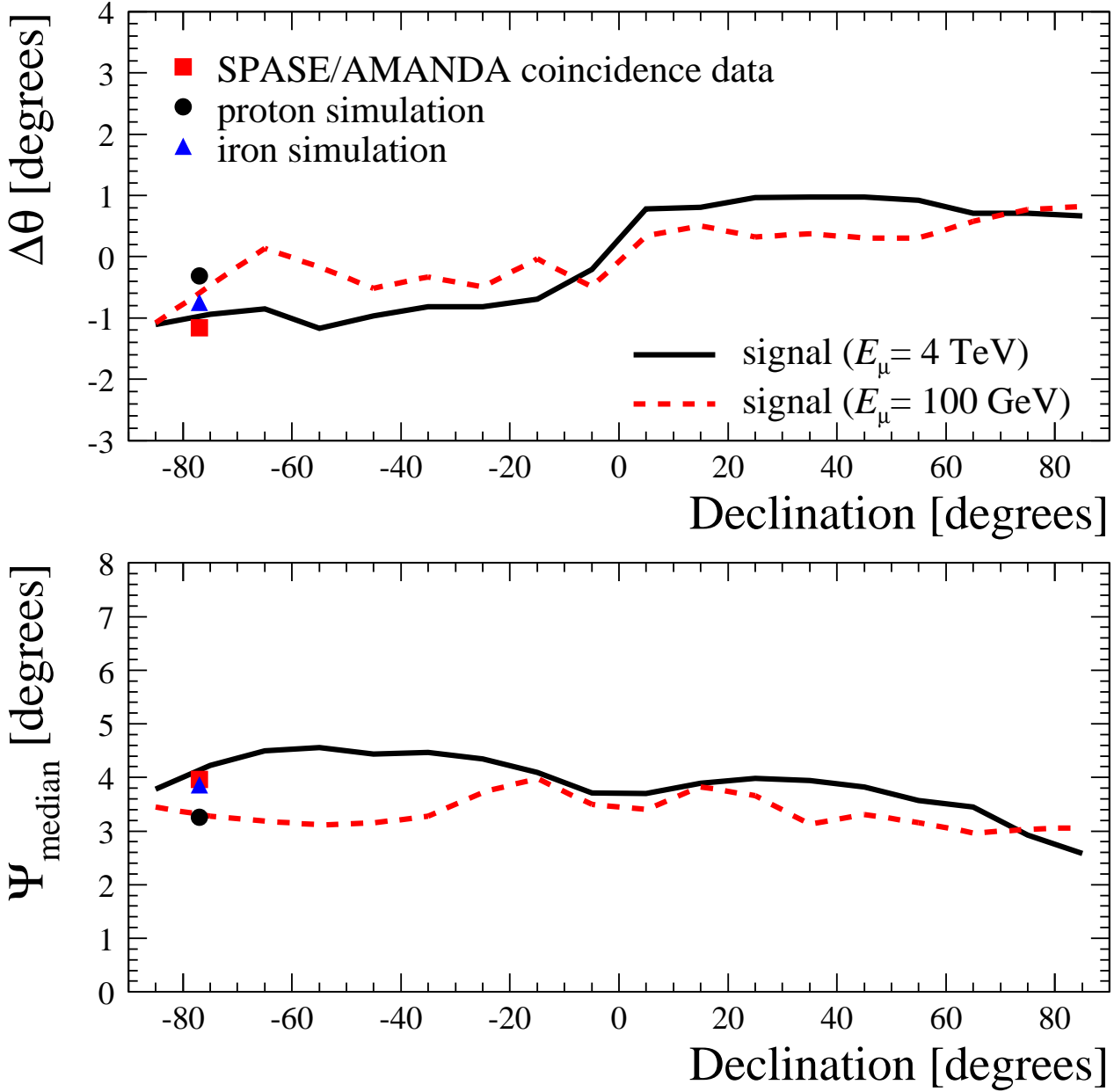}
\caption{Offset in reconstructed zenith angle (upper panel) and median space angle (lower panel) as a 
function of declination.
Positive declination corresponds to upward traveling events in the AMANDA array.
SPASE/AMANDA coincidence data (squares) is compared to expectation assuming that the cosmic ray elemental composition is entirely 
protons (circles) or iron nuclei (triangles). 
Also shown is the expectation for signal (\textit{i.e.}, neutrino-induced muons) with two different energies
within the detector.}
\label{fig:angleres2}
\end{figure}

In this study, AMANDA data was analyzed using the procedure outlined in 
Table~\ref{tb:variables}, with the exception that angle-dependent cuts were inverted to
account for the downgoing direction of travel of SPASE/AMANDA coincidence events (\textit{e.g.},
the cut at stage 3 was changed to $\cos\theta^\mathrm{(5)}>0.1)$.
The absolute pointing accuracy is characterized by the average of $\Delta\theta$,
the difference between the true and the reconstructed zenith angle. Due to the excellent zenith angle resolution of SPASE,
SPASE/AMANDA coincidence data was used to deduce $\Delta\theta$ using the reconstructed zenith angles of both detectors,
$\Delta\theta = \theta_\mathrm{AMANDA}-\theta_\mathrm{SPASE}$. Figure~\ref{fig:angleres2} shows the measured zenith angle resolution using SPASE/AMANDA coincidence events,
together with the resolution obtained for a SPASE/AMANDA simulation of air showers initiated by protons
and iron nuclei. Iron primaries produce a larger fraction of coincidence
events with more than one muon penetrating to AMANDA depths, which accounts for 
the small difference between protons and iron nuclei. 
The coincidence data support the predicted angular resolution, and show that the angular offset is small compared to the angular dimensions of the sky bins. These conclusions are nearly independent of the choice of cosmic ray primary.

Also shown in Fig.~\ref{fig:angleres2} is the expected angular resolution as function of declination
for single muons with energies of 0.1 TeV and 4 TeV within the detector volume.
These muon energies were chosen to be representative of the average muon energy that were initiated
by atmospheric neutrinos and by a source with differential energy spectrum proportional to $E^{-2}$.
The predicted value for the absolute offset is less than 1.5 degrees, which is consistent with results
obtained by additional study of the SPASE\&AMANDA and GASP\&AMANDA coincidence events \citep{Rawlins01,SPAM02}. The offset is due to bias in the AMANDA event reconstruction, which tends to produce more vertical events.  This effect is most evident from the high energy muon simulation which shows that the angular offset changes sign for positive and negative declinations.
Since the absolute offset is substantially smaller than the angular resolution and small compared to the size of 
the search bin, the offset has minimal impact on the signal efficiency.  
For a zenith (declination) offset of 1.5 degrees,  6\% of the signal events are shifted to the neighboring bin.   
The lower panel of the figure also shows that the angular resolution for
upward traveling events is slightly better than for events traveling in
the downward direction, presumably due to the asymmetry in the response of
the photomultiplier tubes, which are oriented toward the center of the earth.  

\begin{figure}
\plotone{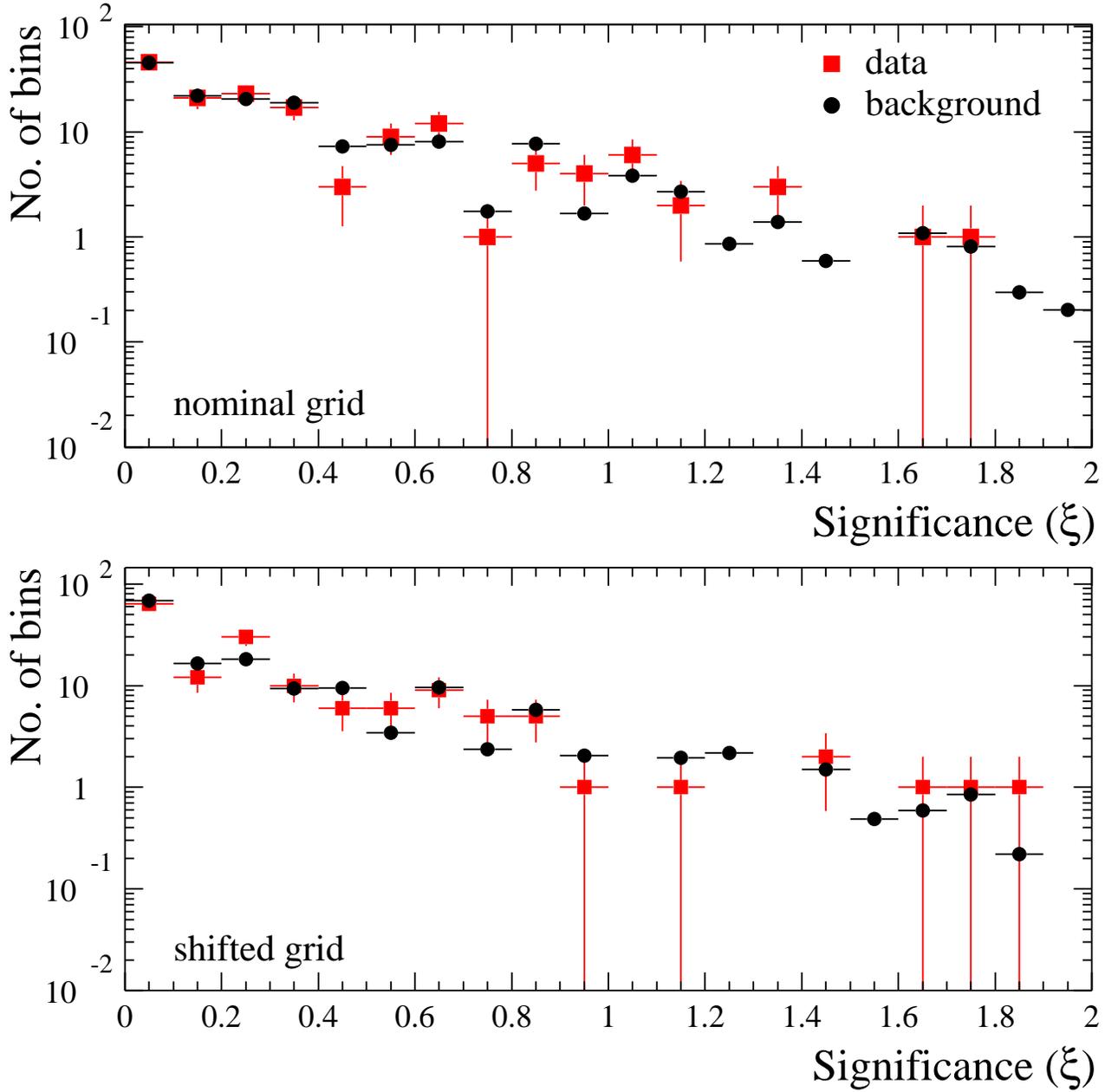}
\caption{Distribution of significance for the 154 sky bins.
Data (squares) is compared to expectation from randomized background (circles). 
The nominal sky grid in the upper panel has been shifted by one half bin in both right ascension 
and declination in the lower panel.}
\label{fig:sigma}
\end{figure}

To obtain approximately equal solid-angle coverage for all bins,
the northern sky is divided into 154 non-overlapping bins, using the calculated optimal declination slice 
($11.25^\circ$) and a varying number of bins in azimuth for the resulting eight declination bands --
from 30 near the horizon to three near $+90^\circ$ declination.
Each angular bin is then tested for an excess of events by computing the significance, 
\begin{equation}
\xi = -\log_{10}(P),
\end{equation}
where 
\begin{equation}
P=\sum_{n=N_{0}}^{\infty} \frac{e^{-\mu} \mu^{n}}{n!} \label{eq:prob}
\end{equation}
is the probability for the bin to contain at least the observed
number of events, $N_{0}$, assuming that fluctuations are 
described by a Poisson distribution.  
The expected mean number of events, $\mu$, is obtained by taking the average of the number
of events in all other bins in the same declination slice. 
The polar location of AMANDA assures equal sky coverage for all
declinations, independent of time gaps in the collection of data.
Figure~\ref{fig:sigma} shows the distribution of significance for the experimental data and for
random fluctuation of the background events.
This noise estimate is obtained by randomizing the right ascension coordinate of the data events,
then recalculating $\xi$ for each bin.
A point source candidate would be identified by a large observed value of significance
with a large ratio to the significance expected from random fluctuations of background.
To avoid the statistical problem of a potential source near a bin boundary distributing signal
between two adjacent bins, the procedure was repeated with the grid shifted by one half bin in 
both declination and azimuth.
The largest value of significance, $\xi = 1.85$, appears in the bottom panel.
Taking into account the 154 bins in the sky and the two versions of the sky grid,
there is a 40\% probability that the most significant sky bin is produced by
random fluctuation of background.
Therefore, the distribution of significance shows no evidence of a source. 

Another approach was also investigated, using the angular correlation function between event pairs to avoid
the problem of a source near a bin boundary, but this alternate approach did not reveal sources either \citep{Young00}.

\section{FLUX LIMITS}
\label{limits}

\begin{figure}
\plotone{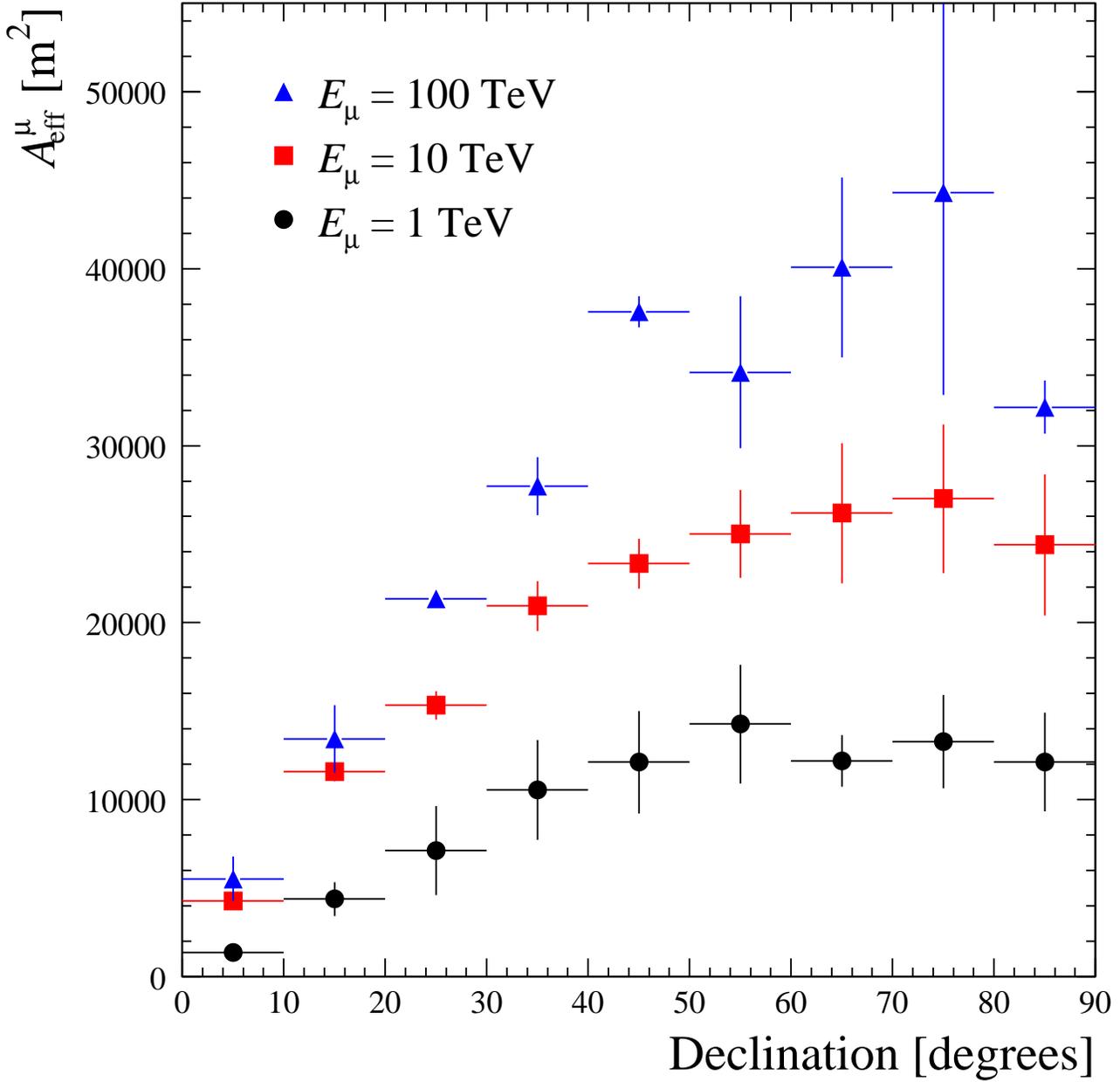}
\caption{AMANDA-B10 effective area for muon detection as a function of declination (90$^{\circ}$ is vertically up)
for three different muon energies at the detector. The vertical error bars are statistical. }
\label{fig:muarea}
\end{figure}

The absence of a detected source translates into an upper limit on the high-energy
neutrino flux.
Neutrino flux and neutrino-induced muon flux limits depend on the effective area of
the detector, $A_\mathrm{eff}$, for a muon with energy $E_{\mu}$.
The effective area, obtained by dividing the signal rate by the incident flux,
is determined from simulations by
\begin{equation}
A_\mathrm{eff}(E_{\mu})= f_\mathrm{ev}(E_{\mu})\cdot A_\mathrm{GEN}
\label{eq:area1}
\end{equation}
where $A_\mathrm{GEN}$ is the cross-sectional area of the cylinder in the 
simulation that contains all neutrino interaction vertices, and $f_\mathrm{ev}$
is the fraction of generated muon events that survive the 13-stage data 
analysis procedure.
As an example,  results are shown in Fig.~\ref{fig:muarea} for muon vertices located near the detector. The effective area is computed as a function of declination for muons with energies of 1, 10 and 100 TeV.
These effective areas can be contrasted to the geometric cross-section of the
instrumented volume of deep ice, which spans from $1.1\times10^4$~m$^2$ in the vertical direction
to $6.1\times10^{4}$~m$^2$ in the horizontal direction.

The \textit{average} effective area, folding in the source spectra, detector response
and the probability of a neutrino-nucleon interaction, is given by 
\begin{equation}
\overline{A}_\mathrm{eff}^{i} = N_{A} \rho_\mathrm{ice} A_\mathrm{GEN} \frac{\int_{E_{\nu}^\mathrm{min}}^{E_{\nu}^\mathrm{max}} \sigma_{\nu \mu} (E_{\nu} ) S(E_{\nu} ) \langle R(E_{\nu} ;E_{\mu}^\mathrm{min} ) \rangle f_{ev} \frac{d\phi_{\nu}}{dE_{\nu}} dE_{\nu}}{\int_{E_{\nu}^\mathrm{min}}^{E_{\nu}^\mathrm{max}} \frac{d\phi_{i}}{dE_{\nu}} dE_{\nu}}
\label{eq:munueffarea}
\end{equation}
where the index $i$ denotes either $\mu$ or $\nu$,
and $E_{\nu}^\mathrm{min}=10$ GeV and $E_{\nu}^\mathrm{max}=10^7$ GeV
define the energy range in the simulation, which safely brackets the interval
of interest for most theoretical models.
The other variables are $N_{A}$, Avogadro's number, $\rho_\mathrm{ice}$, the molar density of nucleons in ice, and $\sigma_{\nu \mu}$ is the charged-current cross section
for $\nu_{\mu}$ (or $\overline{\nu}_{\mu}$) interactions.
The term $S(E_{\nu})$ accounts for neutrino absorption in the earth and 
$\langle R(E_{\nu} ;E_{\mu}^\mathrm{min} ) \rangle$ is the average
propagation length for a muon created by a neutrino with energy $E_{\nu}$, corrected for the
energy threshold of the detector.
In this calculation, the neutrino interaction vertices are located randomly
within a volume that is large compared to the instrumented volume of the detector.  The propagation programs properly account for muon energy losses.

\begin{figure}
\plotone{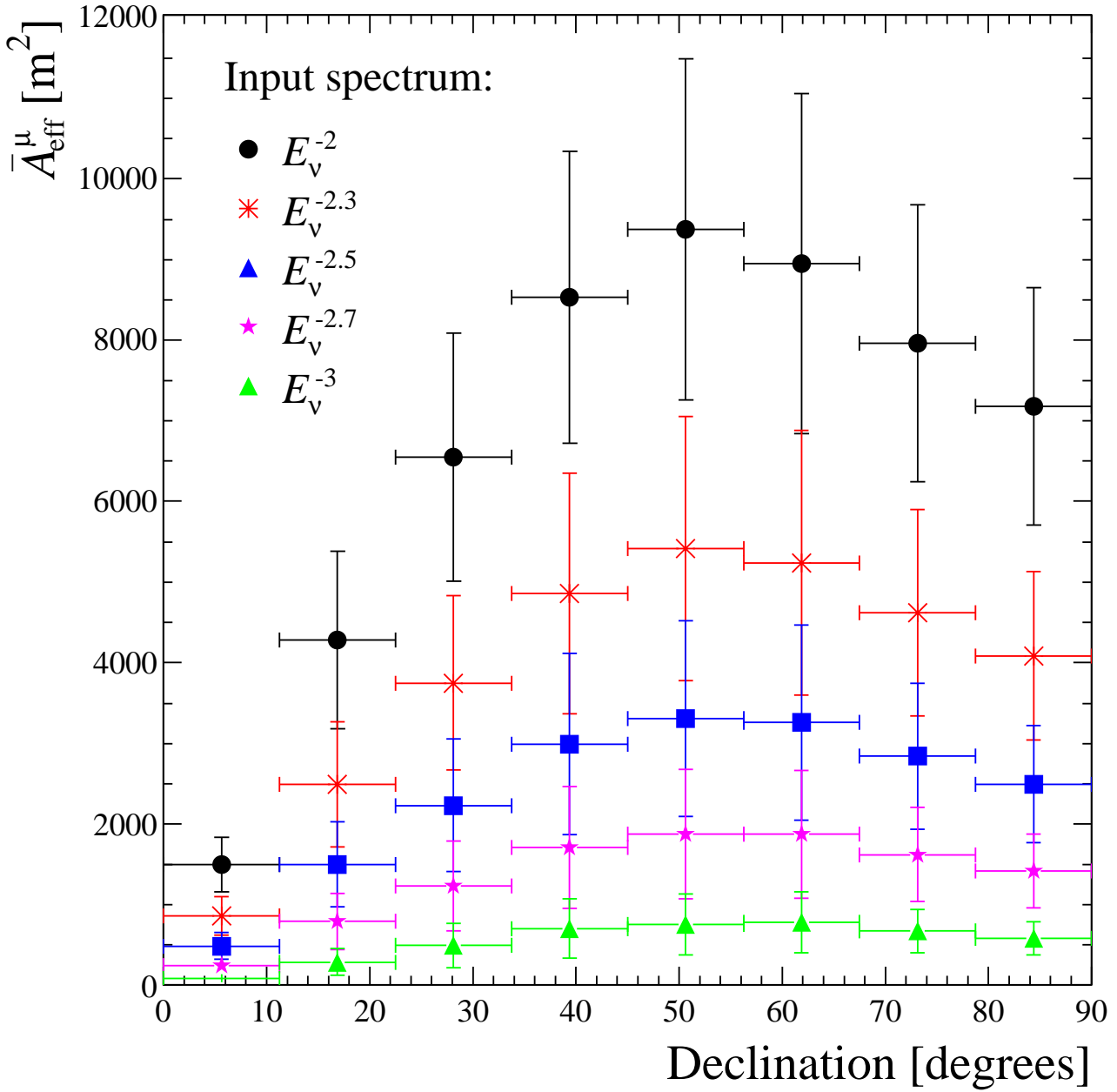}
\caption{AMANDA-B10 average effective area for muon detection as a function of declination (90$^{\circ}$ is
vertically up) for assumed differential neutrino spectral indices between 2 and 3.
The vertical error bars indicate the uncertainty obtained from studies described in Sec.~\ref{systematics}. }
\label{fig:avemuarea}
\end{figure}

\begin{figure}
\plotone{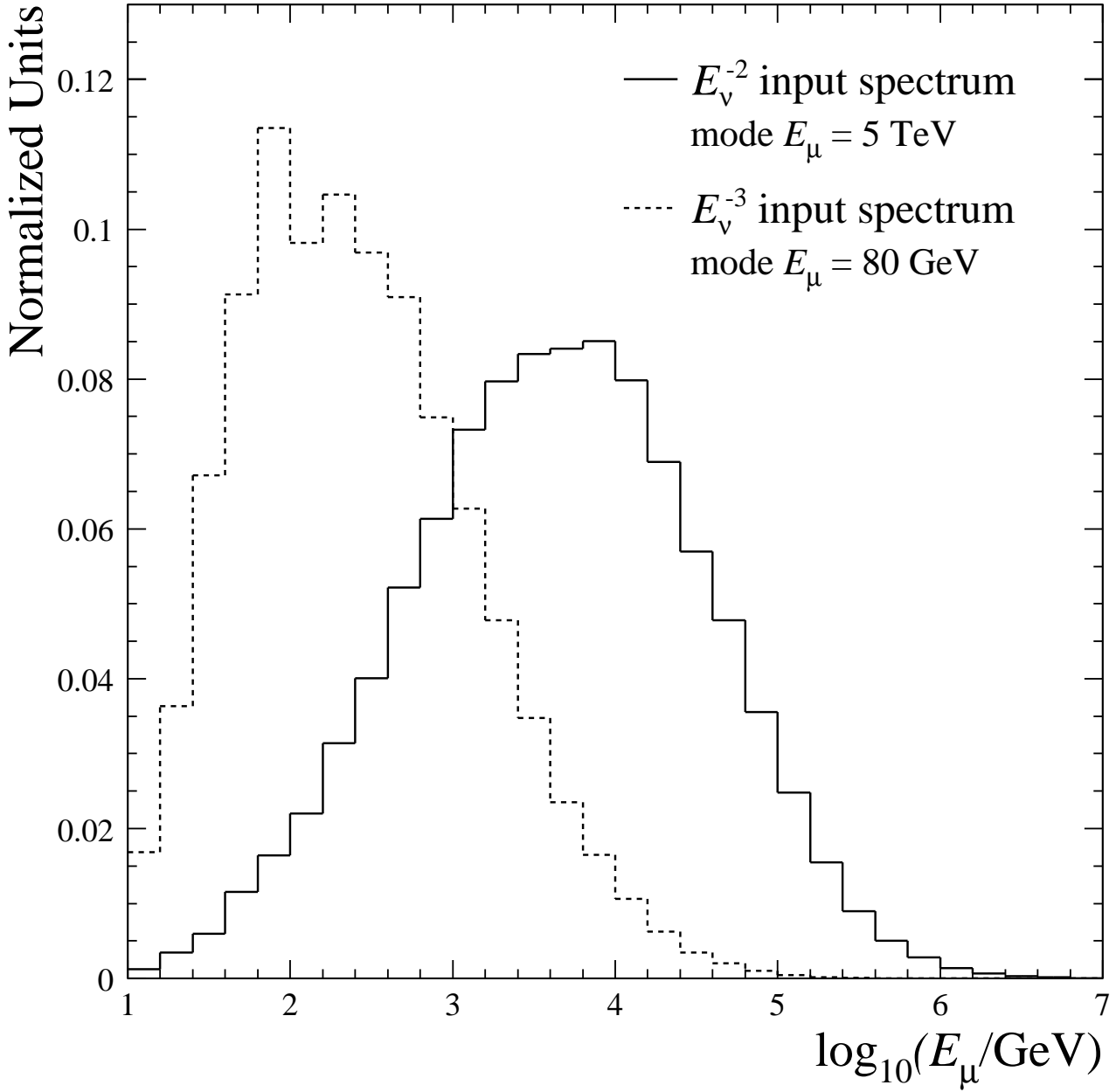}
\caption{Neutrino-induced muon energy distribution at the detector. 
The differential neutrino spectra used as input for the simulation are proportional to $E^{-2}$
(solid line) and $E^{-3}$ (dashed line) respectively. The most probable energy for each distribution is shown. }
\label{fig:mu_energy}
\end{figure}

The muon differential flux is related to the neutrino differential flux by
\begin{equation}
\frac{d\phi_{\mu}}{dE_{\nu}}=\sigma_{\nu \mu} (E_{\nu} ) S(E_{\nu} ) \langle R(E_{\nu} ;E_{\mu}^\mathrm{min} ) \rangle \frac{d\phi_{\nu}}{dE_{\nu}} \, . \label{eq:muondifflux}
\end{equation}
The energy-averaged muon effective area is presented in Fig.~\ref{fig:avemuarea} for different spectral indices. 
The energy response of AMANDA is described by the distribution of 
$E_{\mu}$ at the detector, which depends on the spectral index of
the neutrino spectrum.
Figure~\ref{fig:mu_energy} shows the distribution for  differential
spectra proportional to $E^{-2}$ and $E^{-3}$.
For $E^{-2}$, the most probable muon energy (mode) is 5 TeV, and 
the central 90\% of the muon events are within the energy interval
80 GeV to 200 TeV.
For a spectral index of 3, appropriate for atmospheric neutrinos,
the most probable detected energy of the muon is much lower.

\begin{figure}
\plotone{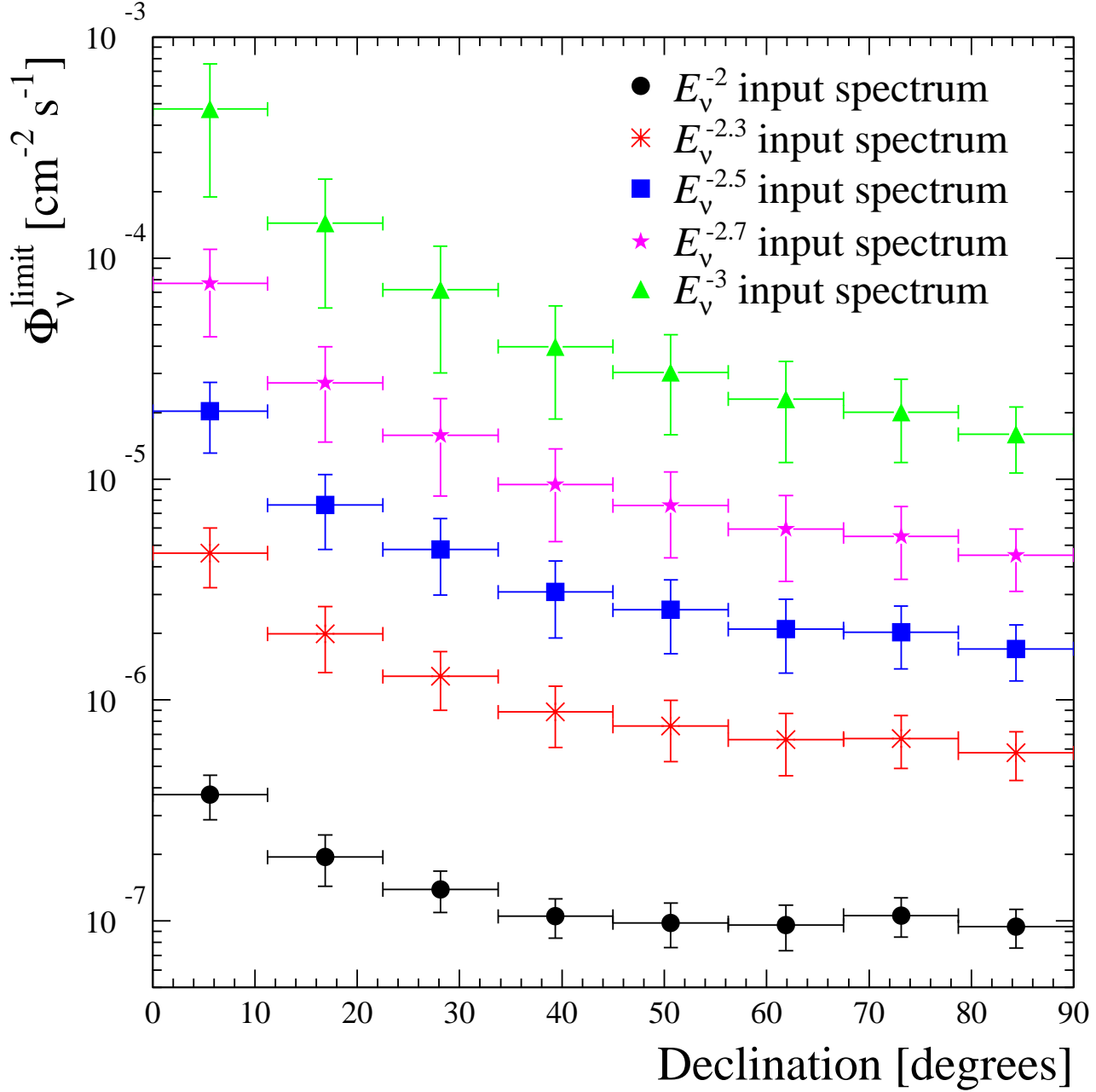}
\caption{Neutrino flux limit (90\% CL) for various spectral indices.
The results are shown as a function of declination, averaged over right ascension.
Power law exponent refers to the differential neutrino energy spectrum.
The vertical error bars indicate the uncertainty obtained from systematic studies.}
\label{fig:nulimit}
\end{figure}

\begin{figure}
\plotone{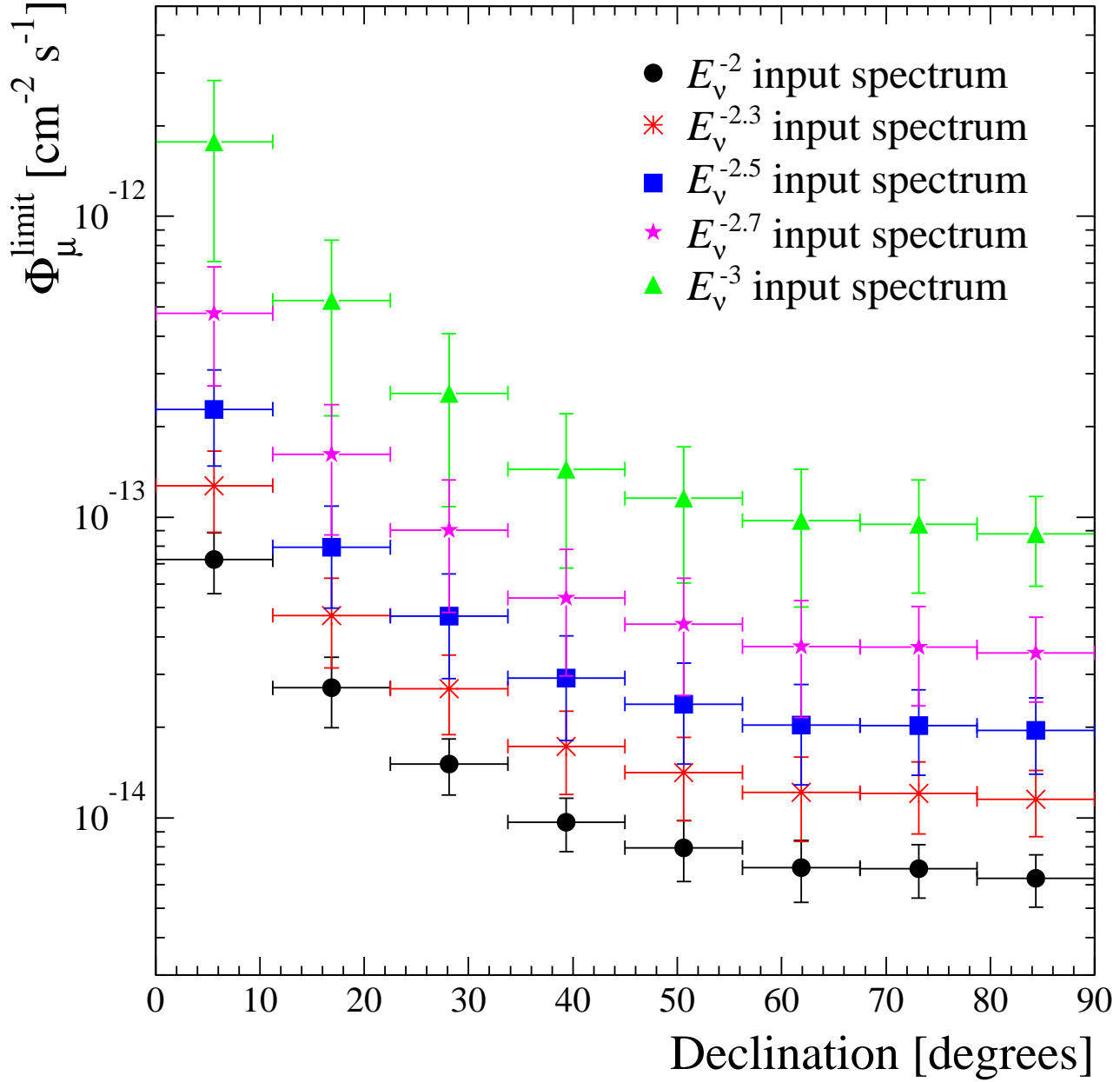}
\caption{Neutrino-induced muon flux limit (90\% CL) for various spectral indices.
The results are shown as a function of declination, averaged over right ascension.
Note that the power law exponent refers to the differential neutrino energy spectrum.}
\label{fig:mulimit}
\end{figure}

If neutrino emission from a point source is described by a differential energy
spectrum, the energy-averaged flux limit is calculated from
\begin{equation}
\Phi_{i}^\mathrm{limit}  =  \frac{\mu_\mathrm{s} (N_{0} , N_\mathrm{bgr} ) }
{T_\mathrm{live} \cdot \epsilon_\mathrm{bin} \cdot \overline{A}_\mathrm{eff}^{i}}\;. \label{eq:fluxlimit2}
\end{equation}
The quantity $\mu_\mathrm{s} (N_{0} , N_\mathrm{bgr})$ is the upper limit on
the number of signal events at 90\% confidence level, calculated following the
unified procedure of \citet{stat},
where $N_0$ is the observed number of events in a potential source bin and $N_\mathrm{bgr}$ is
the expected number of background events.
For the binning technique employed in our point source search,
$N_\mathrm{bgr}$ is determined by averaging the number of observed
events over the declination band, excluding the bin being considered.
The efficiency factor, $\epsilon_\mathrm{bin}$, accounts for the finite
angular resolution and the possible noncentral location within the bin
of a potential source.
The factor $T_\mathrm{live}$ is the operational livetime of the detector.
The resulting muon and neutrino flux limits are shown in Fig.~\ref{fig:nulimit}
and Fig.~\ref{fig:mulimit}, respectively, for various assumed spectral indices between 2 and 3. 

\begin{figure}
\plotone{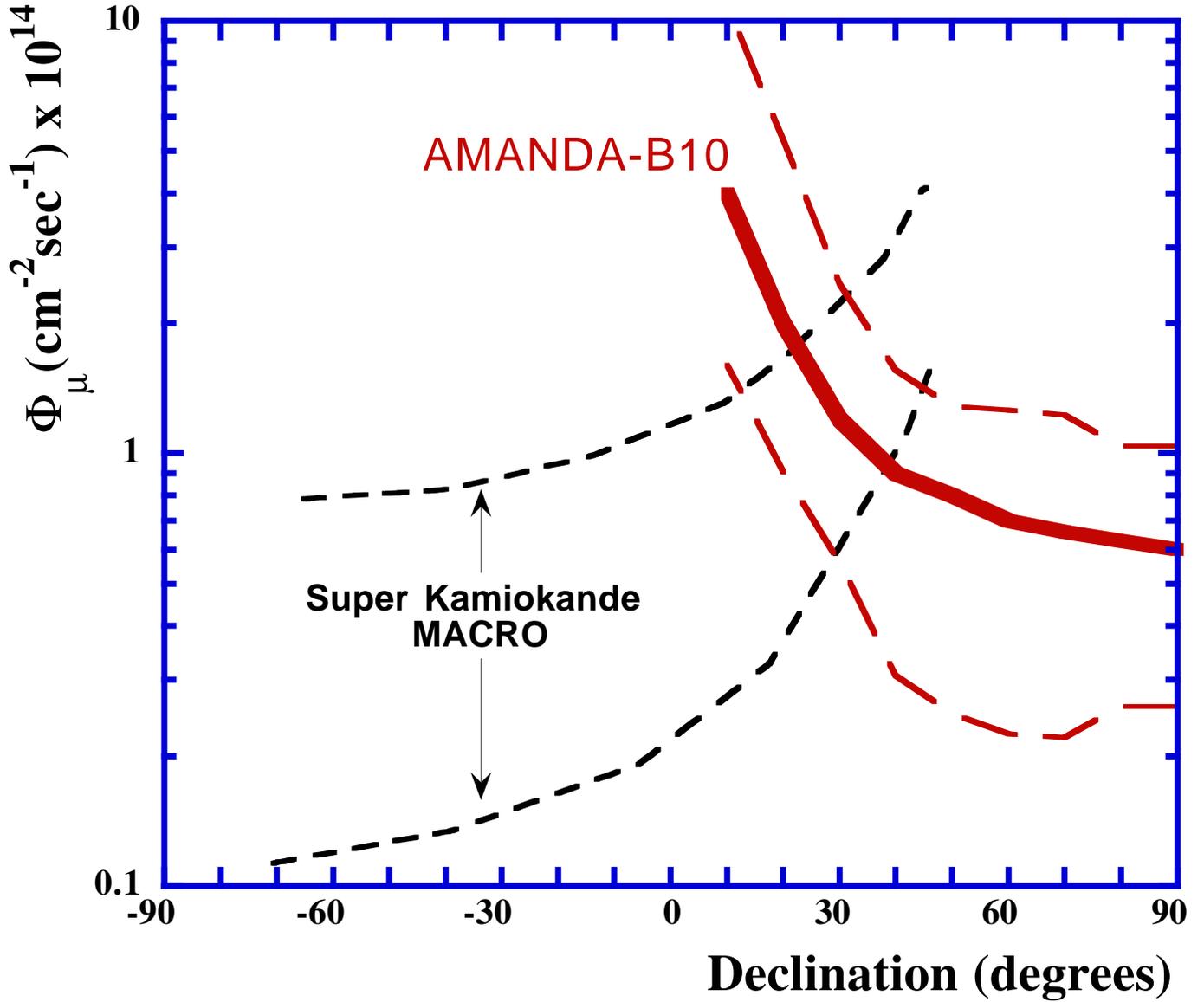}
\caption{Upper limit on the muon flux (90\% CL) as a function of declination.
The solid curve is the AMANDA-B10 limit, averaged over right ascension, and the region delineated
by the long dashed curves provides a guide to the statistical fluctuation within the declination interval (see Table~\ref{tb:allskyflux}).
The band defined by the short dashed lines indicate the range of limits presented by MACRO \citep{MACRO99}
and Super Kamiokande \citep{SK01}.  }
\label{fig:global}
\end{figure}

Figure~\ref{fig:global} shows a comparison of the AMANDA flux limits with a
representative sample of neutrino detectors located in the Northern Hemisphere.
The AMANDA-B10 detector approaches its maximum sensitivity for declinations
greater than +30$^{\circ}$, which complements the sky regions covered by
neutrino detectors such as MACRO \citep{MACRO99} and Super Kamiokande \citep{SK01}.
With only 130.1 days of detector livetime, the muon flux limits for positive
declinations approach those achieved for the Southern sky. 
The AMANDA flux limits were calculated for $E_{\nu}>10$~GeV, 
while both Super Kamiokande and MACRO present fluxes for $E_{\mu}>$ 1-2 GeV,
but for relatively hard differential neutrino spectra, such as $E_{\nu}^{-2}$,
the impact of energy threshold on muon flux limits is modest \citep{Biron02}.  

\begin{figure} 
\plotone{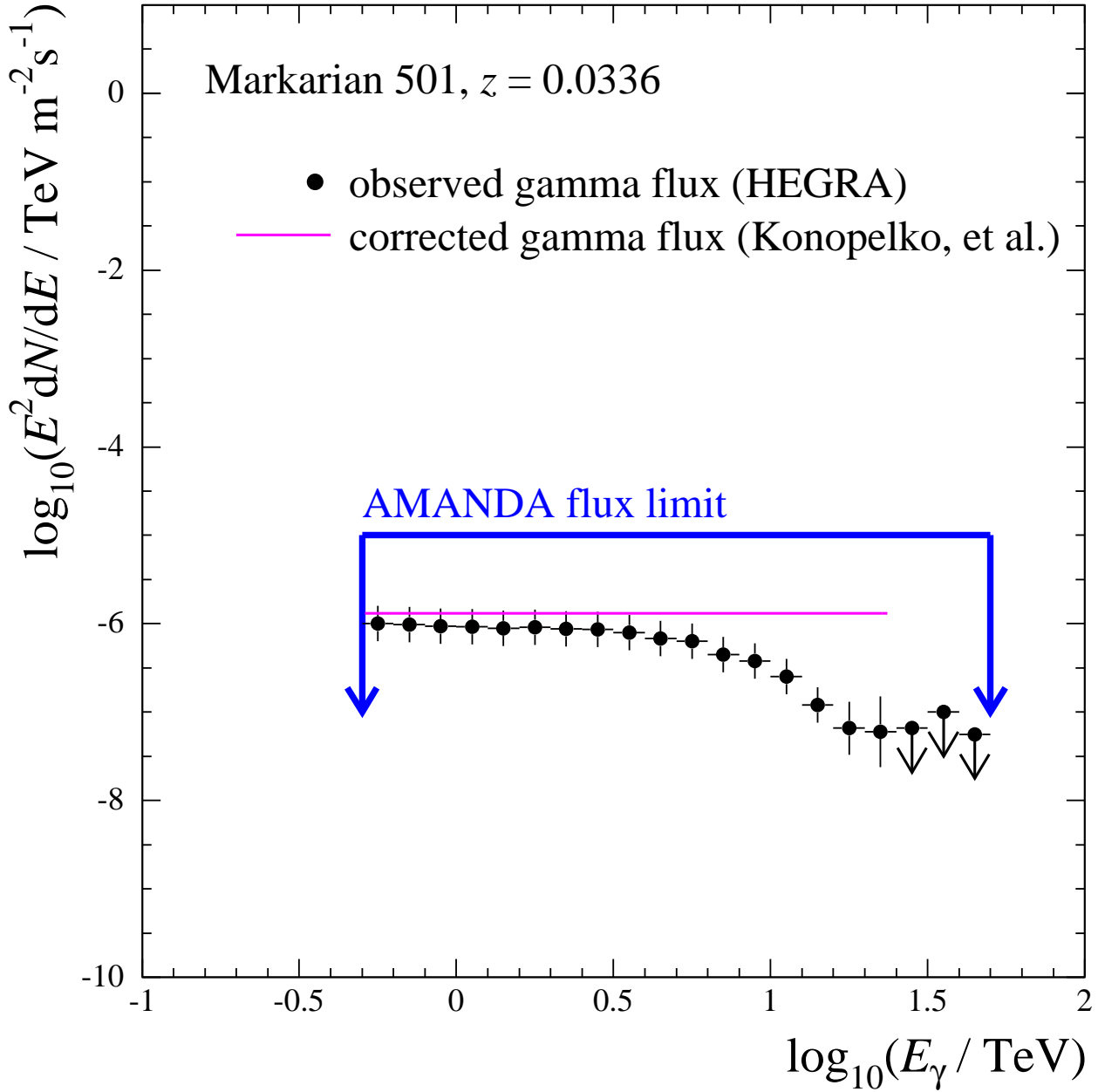}
\caption{Time averaged spectrum of gamma rays from Markarian 501 observed in 1997 \citep{HEGRA1997,Whipple501}
and corrected for intergalactic absorption by the diffuse infrared background \citep{Konopelko99}.
Gamma flux is compared to AMANDA neutrino limit assuming an energy dependence on the neutrino flux proportional
to $E^{-2}$.}
\label{fig:HEGRA501}
\end{figure}

In addition to the general search for a point source, a number of potential sources of particular
interest were investigated by performing the significance test while centering the search bin on
their sky coordinates.
The resulting flux limits are presented in Table~\ref{table:limit}.
As one interesting example, we compare the AMANDA limit on the neutrino flux from
Markarian 501 to the observed gamma ray flux and this flux corrected for intergalactic
absorption by infrared photons in Fig.~\ref{fig:HEGRA501}.
Assuming that the neutrino energy spectrum is proportional to the inferred gamma-ray 
spectrum at the source, the AMANDA limit constrains the proportionality factor.
For example, the ratio of the flux of neutrinos to the flux of gamma-rays must be less than ten if the source spectrum of
\citet{Konopelko99} is assumed. Recent work \citep{DeJager02} suggests that the ratio may be smaller.

\section{IMPACT OF SYSTEMATIC UNCERTAINTIES ON FLUX LIMITS}
\label{systematics}

In the absence of a well-understood source of high-energy neutrinos, the sensitivity of
the AMANDA-B10 detector, as expressed in terms of the effective area and angular resolution, had to be estimated from 
detector simulations.
The required input relies on knowledge of detector performance extracted from, \textit{e.g.},
laboratory measurements of the individual components, \textit{in-situ} measurements of the optical
properties of the ice, and calibration studies.
Consequently, the predicted sensitivity is affected by uncertainty in this information.
Table~\ref{tb:systematics} lists the dominant contributions to systematic uncertainties.
The uncertainty in the right column is defined as the variation 
$\arrowvert (A_\mathrm{eff}^{j}-A_\mathrm{eff}^\mathrm{nom}) /
(A_\mathrm{eff}^{j}+A_\mathrm{eff}^\mathrm{nom})\arrowvert$ 
of the effective area from its value determined with the nominal set of input parameters,
$A_\mathrm{eff}^\mathrm{nom}$,
given by the area, $A_\mathrm{eff}^{j}$,
obtained by varying the specified parameter (index $j$) by its estimated uncertainty.

The most significant component is generated from the uncertainty in the angular dependence of the OM sensitivity.
It arises mainly from a lack of detailed understanding of the physics governing the refreezing process
in the water column required to be melted for the deployment of OMs.
A local increase in scattering from air bubbles trapped in the vicinity of the OM 
translates into a modulation of its angle-dependent acceptance.
This effect is difficult to disentangle from the intrinsic angular dependence of
the OM sensitivity, which was measured in the laboratory.
An event sample highly enriched in atmospheric muons was used to investigate the \textit{in situ} angle dependence
of the OM sensitivity \citep{atm_nu02}.
The modification to the angular sensitivity leads to a 25\% uncertainty in the effective area.

Since the angle-integrated sensitivity of the OM is a poorly constrained parameter in this analysis,
we also investigated the impact of varying the absolute sensitivity of the OM.
It was parameterized by a wavelength-dependent function that included the PMT quantum efficiency,
the OM collection efficiency, obscuration by nearby cables, and absorption properties of
the glass pressure vessel and coupling gel.
We obtain a fractional uncertainty of 0.15 in the effective area after reducing the absolute OM sensitivity
by 15\%, a value consistent with the atmospheric neutrino results.
Further reduction is inconsistent with observed experimental trigger rates.

As mentioned in Sec.~\ref{simulation}, two muon propagation routines were employed to show that
systematic variations in the effective area for signal (\textit{i.e.,} upward traveling) muons were between 5\% and 10\%.  This is much less than observed for studies of atmospheric muons, presumably due to the much weaker angular dependence of the average pathlength.
Possible uncertainty in timing and position calibration of individual OMs are included
by varying these parameters to the largest extent allowed by the imprecision of the calibration procedures.
The effective area changes by 10\%. 
A conservative estimate of the variation in sensitivity introduced by uncertainties in the depth-dependent optical
properties and their approximate treatment in the detector simulation is obtained by substituting
the nominal bulk ice model, containing a parameterization of the measured dust strata,
with a homogeneous ice model.
The impact on effective area is less than 5\%.

\begin{figure} 
\plotone{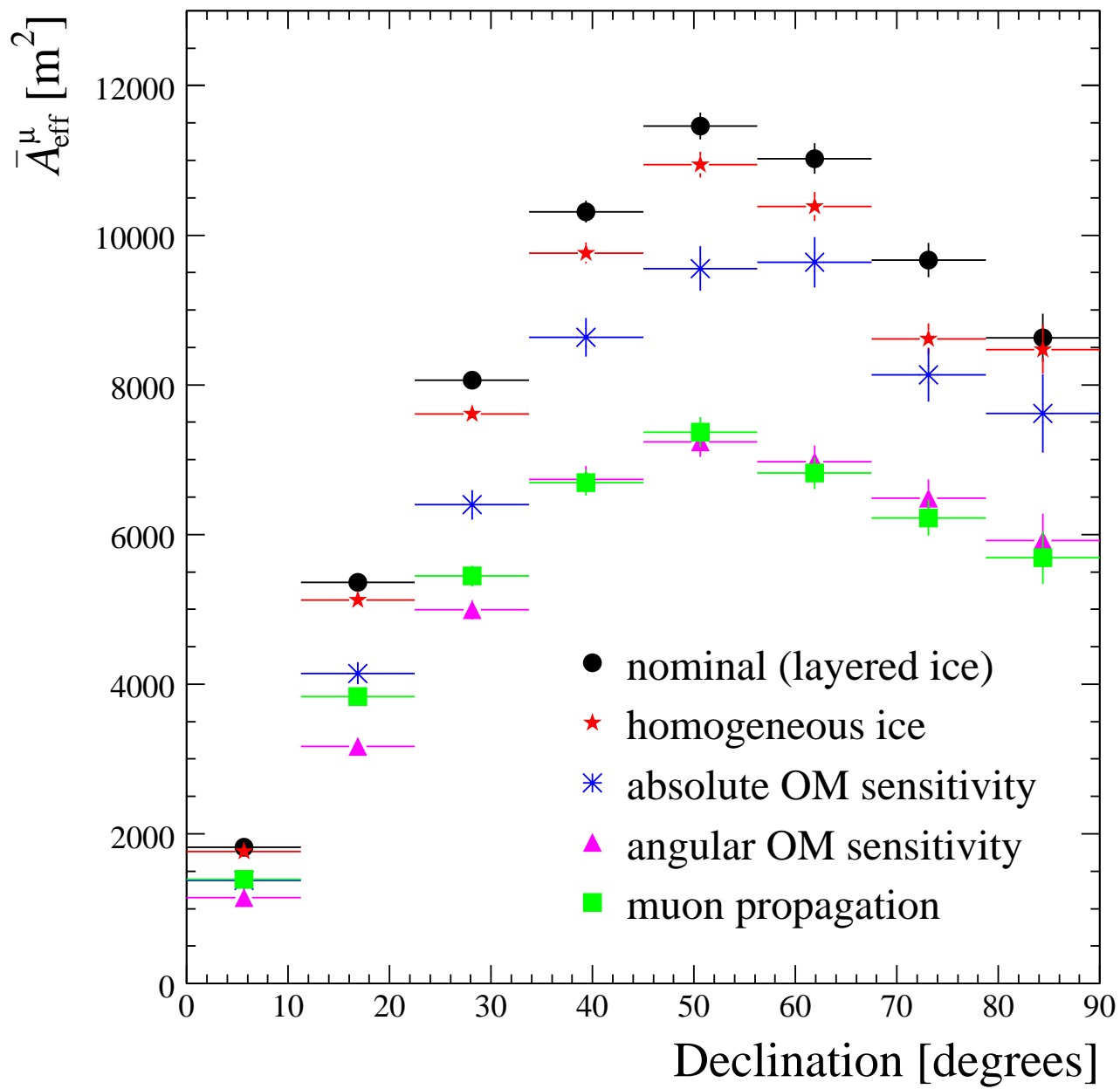}
\caption{Comparison of average muon effective area for differential neutrino signal proportional
to $E^{-2}$ . See text for explanation of legend.}
\label{fig:area_sys}
\end{figure}

The impact of the most dominant systematic uncertainties on the average effective area is shown
in Fig.~\ref{fig:area_sys}, where the systematic variations have been applied one at a time,
with the exception of the muon propagation curve which also includes the variation of the
angular OM sensitivity.

\begin{figure} 
\plotone{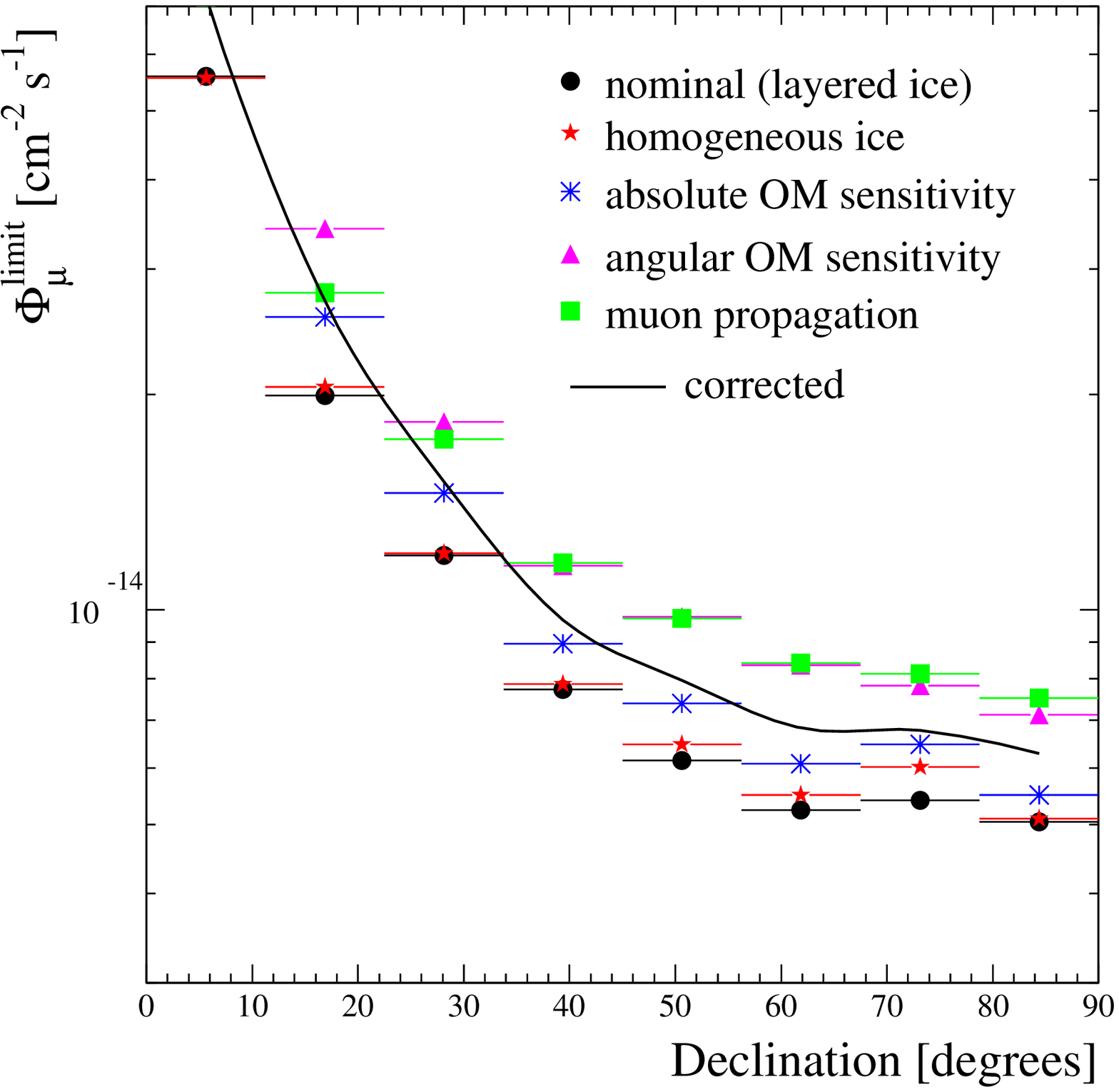}
\caption{Comparison of muon flux calculations for differential neutrino signal proportional to $E^{-2}$.
See text for explanation of legend. The solid line (corrected) includes systematic uncertainty and indicates the final result of this work.}
\label{fig:muflux_sys}
\end{figure}

\begin{figure} 
\plotone{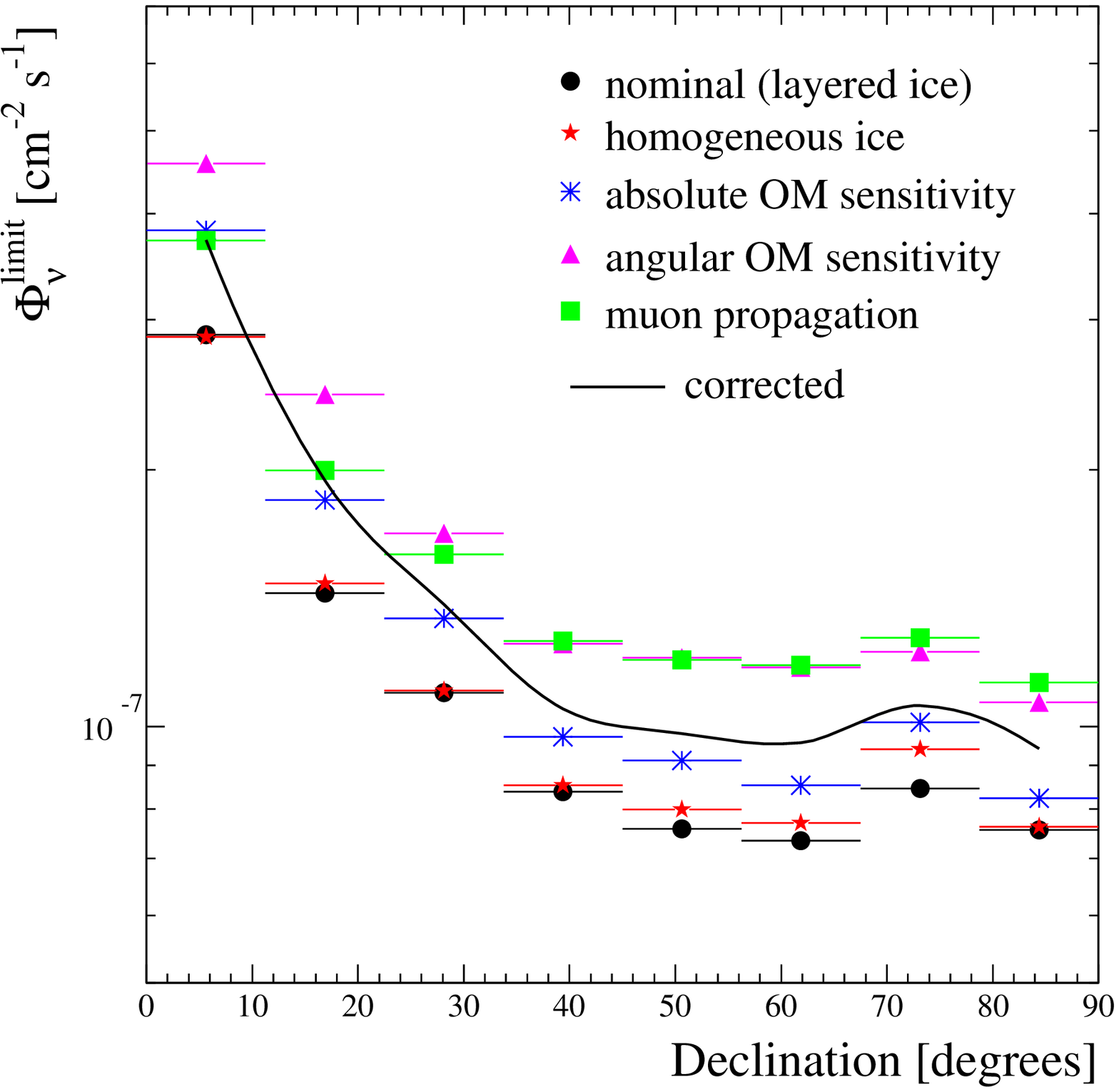}
\caption{Comparison of neutrino flux calculations for differential neutrino signal proportional to $E^{-2}$.
See text for explanation of legend. The solid line (corrected) includes systematic uncertainty and indicates the final result of this work.}
\label{fig:nuflux_sys}
\end{figure}

The variation of the detector sensitivity due to systematic uncertainties was studied by adjusting the physical parameters in the detector simulation. The parameters were adjusted according to the known or estimated uncertainties listed in Table~\ref{tb:systematics}, which are assumed to bound the true values of the parameters.  Systematic uncertainty were included according to the prescription of \citet{Conrad02} which is an extension of the method of \citet{Cousins92}. The calculations assumed that the distribution of systematic uncertainty was flat, and bound by the maximum and minimum values for a given declination bin found in 
Fig.~\ref{fig:muflux_sys} and Fig.~\ref{fig:nuflux_sys}.

The solid curves in Fig.~\ref{fig:muflux_sys} and Fig.~\ref{fig:nuflux_sys} indicate the flux limits after adjusting for systematic uncertainty. They are valid for declination greater than +5$^{\circ}$.
The limits including systematic uncertainties are about 25\% worse than those obtained from the
simulation with nominal input parameters. Finally, the flux limits change by less than 6\% due to the effects of zenith offset and the variation
in $\Psi_\mathrm{median}$ due to declination \citep{Young00}. These small effects were not taken into consideration in the limit calculations.

\section{DISCUSSION}
\label{discussion}

The previous sections have shown that AMANDA-B10 has unprecedented sensitivity to high energy neutrinos and possesses the necessary angular response and background rejection to search for point emission of these particles from astronomical objects; \textit{i.e.,} it is a novel telescope that detects the neutrino messenger. The sensitivity and angular response were determined by simulation.  The reliability of these programs was established by utilizing the known signals generated by (downgoing) atmospheric muons and (upgoing) atmospheric neutrinos.  The angular response was confirmed by the study of air shower events that triggered both AMANDA-B10 and SPASE. Systematic uncertainty in the analysis procedure was also addressed.  

The search for point sources of high-energy neutrinos revealed no candidates. 
A set of event selection criteria was determined by optimizing the signal to 
noise ratio for a signal with a hard energy spectrum, yet this analysis retains reasonable sensitivity for softer spectra.
The upper limits on muon flux for all search bins in the northern hemisphere
are presented in Table~\ref{tb:allskyflux}.

The neutrino flux limits in Fig.~\ref{fig:nulimit} are inferred from the assumption of a power-law
energy spectrum.
This procedure is reliable if the mean energy of the neutrino-induced muon is compatible with the
energy response of the detector.
For example, Fig.~\ref{fig:mu_energy} shows that $E_{\mu}$ at the detector brackets the interval
between 0.1~TeV and $10^3$~TeV for source spectra proportional to $E^{-2}$.
Two lines of evidence show that the simulated energy response of the detector is valid
over this interval.
First, the agreement between the detected and expected rates of atmospheric neutrinos shows that
the response of AMANDA is being correctly modeled in the sub-TeV region.
Second, the tails of the \nch\ distribution are sensitive to brighter events within
AMANDA, which are roughly equivalent to single muons with energy above 1~TeV.
We know of no reason to doubt the predicted energy response for $E_{\mu}<10^{3}$ TeV.
Evaluation and calibration of the energy response beyond $10^{3}$ TeV remain an ongoing activity. 

Not all model predictions are well characterized by power-law energy spectra.
Therefore, Table \ref{table:limit} shows the results for a selection of models
in the literature.
The inferred limits on neutrino flux apply to point sources with continuous 
emission (or episodic emission averaged over the time interval of data collection)
and power-law energy spectra with a fixed spectral index. 
The limits presented here for sources at large positive declination complement
existing data, so that comparable limits now exist for the entire sky.

During 1997, the TeV gamma-ray emission of two nearby AGN blazars (Markarian 421 and 501) were observed to
exhibit episodic flaring. If neutrino emission follows the same time variability, then it may be possible
to improve the signal-to-noise ratio by eliminating the periods of relatively low output.
Multiple detection of Mkn~501 from several air Cherenkov instruments allowed nearly continuous monitoring,
including periods when the moon was shining.
However, monitoring by multiple instruments only extended from March to late August.
Due to uncertainties in the details of the time dependence of the gamma emission, \textit{neutrino}
flux limits are not greatly improved by restricting the analysis to high-flux periods of gamma-ray emission. 

While this paper describes an analysis dedicated to the search for point sources, another strategy was developed based on the event selection of the atmospheric neutrino analysis \citep{Biron02}. The results of this complementary analysis are consistent with the results presented here. The absolute efficiency was extracted by comparing to the known flux from atmospheric neutrinos. Moreover, the second analysis was subject to different systematic uncertainties.

\begin{figure} 
\plotone{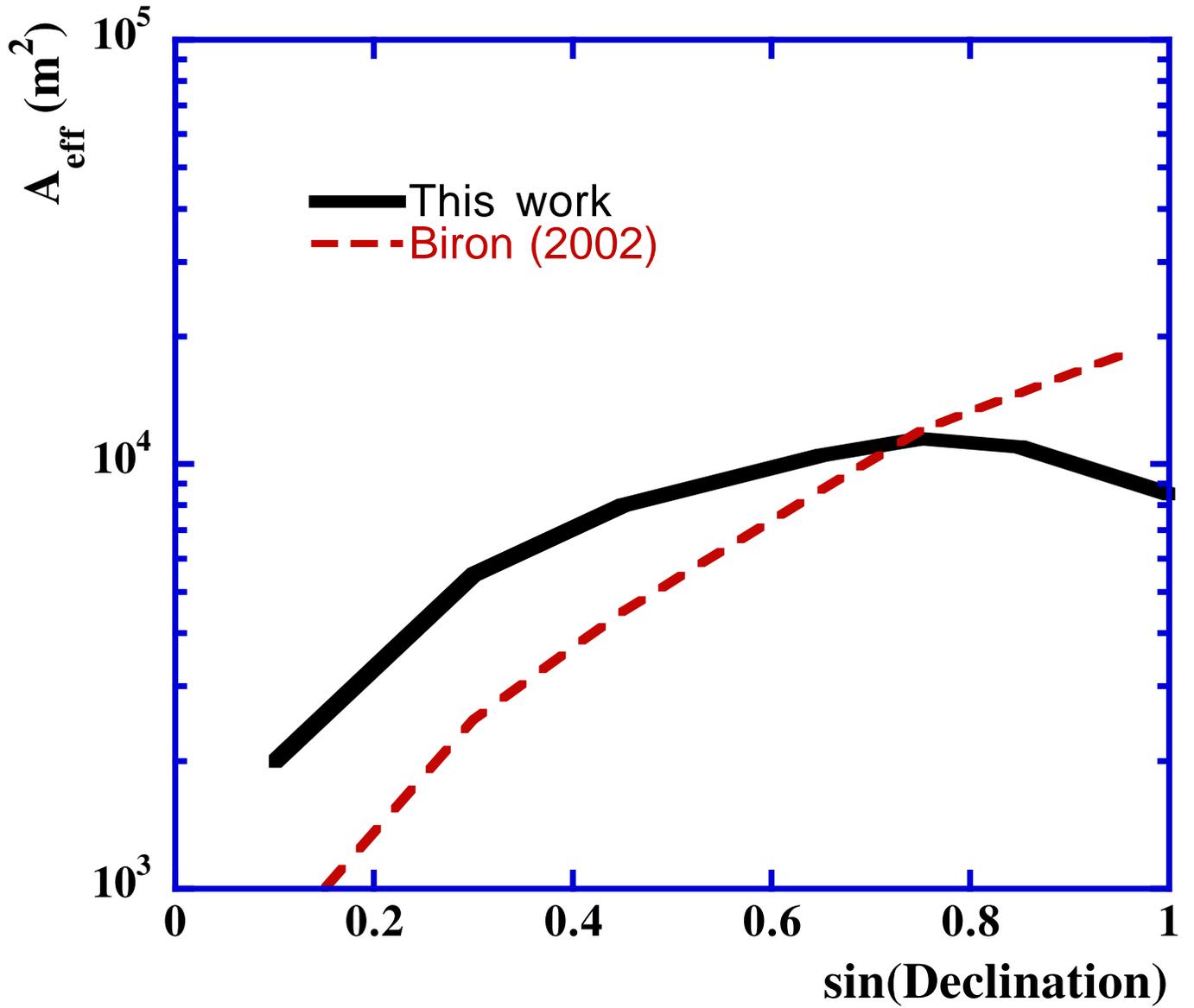}
\caption{ Comparison of average muon effective area as a function of declinations for differential neutrino signal proportional to $E^{-2}$. The dashed curve is from \citet{Biron02}. Neither curve includes the effect of systematic uncertainties.}
\label{fig:biron_comp}
\end{figure}

The method based on the atmospheric neutrino analysis retained a smaller event sample of 369 events, of which $\sim270$ are expected from atmospheric neutrinos. The cut selections produce an implicit optimization on more vertical events and/or softer energy spectra. Figure~\ref{fig:biron_comp} compares the average effective area of the two analysis for an assumed differential spectra proportional to $E^{-2}$. The best flux limits for soft spectra are obtained by atmospheric neutrino analysis, but the neutrino and muon flux limits for either analysis are much larger than obtained for an assumed power law of index of -2.0. 

While the flux limits for any particular source or direction in the northern hemisphere can be extracted from this analysis (see Table~\ref{tb:allskyflux}), flux limits -both integral and pseudo-differential- for a pre-selected list of 62 sources have been reported \citep{Biron02}.  These include all known TeV gamma ray blazars, nearby QSOs, and galactic TeV gamma ray sources in the northern hemisphere. The list also includes microquasars, the five most luminous AGNs in wavelength bands that span across MeV, X-ray, infrared, and radio bands. Of particular interest are radio galaxies with strong emission at GHz frequencies. We have also investigated BL Lacs that are close to the arrival directions of the very highest energy cosmic rays \citep{Tinyakov} and the 10 reported cosmic ray doublets at extreme energies \citep{Uchihori}.

\section{FUTURE}
\label{future}

The technique employed in this paper optimized the selection criteria on signal to noise. Due to the relatively large number of sky bins and the relatively low 
number of events in any individual bin, the analysis procedure produced event
samples that are dominated by poorly-reconstructed atmospheric muons rather
than upward traveling atmospheric neutrino background.  However, as the background rejection of downgoing events improves with larger detectors, such as AMANDA-II, this trend may not continue. 

AMANDA-II, completed in January 2000, surrounds the B10 core with nine 
additional strings, more than doubling the number of optical modules.
For this broader configuration, the effective area for neutrino-induced muons 
remains relatively constant over the entire hemisphere \citep{Barwick01}.
Consequently, AMANDA-II is expected to achieve a factor of
five improvement in sensitivity for nearly horizontal events compared to AMANDA-B10 \citep{Wischnew01}.  The greater statistical sample of atmospheric neutrinos will allow better tests of the detector simulation programs, especially near the horizon. 
With the data already collected on tape, AMANDA-II can observe (or exclude) neutrino fluxes that are approximately one order of magnitude below the limits presented here, as shown in Fig.\ref{fig:pointlim}.

\section{ACKNOWLEDGEMENTS}
This research was supported by the following agencies:
U.S. National Science Foundation Office of Polar Programs
and Physics Division,
University of Wisconsin Alumni Research Foundation,
U.S. Department of Energy,
Swedish Natural Science Research Council,
Swedish Polar Research Secretariat,
Knut and Alice Wallenberg Foundation, Sweden,
German Ministry for Education and Research, 
the U.S. National Energy Research Scientific Computing Center 
(supported by the Office of Energy Research of the U.S. Department of Energy),
UC-Irvine AENEAS Supercomputer Facility,
Deutsche Forschungsgemeinschaft (DFG).
C.~P\'erez de los Heros received support from the EU 4th framework of Training
and Mobility of Researchers, and D.~F.~Cowen acknowledges the support of the
NSF CAREER program. P.~Desiati\ was supported by the Koerber Foundation.

\clearpage
\newpage 

\clearpage
\newpage

\clearpage
\newpage

\clearpage
\newpage

\clearpage
\newpage

\clearpage
\newpage

\clearpage
\newpage

\clearpage
\newpage

\clearpage
\newpage

\clearpage
\newpage

\clearpage
\newpage

\clearpage
\newpage

\clearpage
\newpage

\clearpage
\newpage

\clearpage
\newpage

\clearpage
\newpage

\clearpage
\newpage

\clearpage
\newpage

\clearpage
\newpage

\clearpage
\newpage

\clearpage
\newpage

\clearpage
\newpage

\clearpage
\newpage
\begin{table*}
\caption{Description of Variables Used in Analysis\label{tb:variables}} 


\footnotesize
\begin{tabular}{cllccccc}  
\tableline
\tableline
 Stage&Selection Cut & $\epsilon_\mathrm{data}$ & $\epsilon_\mathrm{bgr}$ & $\epsilon_\mathrm{sig}$ & $R_\mathrm{data}$ & $R_\mathrm{bgr}$ \\ \tableline
 0 & Trigger 
  & 1 & 1 & 1 & --- & --- \\ 
 1 & Filter 1: 
  & 0.0193 & 0.0190 & 0.433 & 22.4 & 22.8 \\
 \multicolumn{1}{r}{ a} & $\theta^{(1)} > 50^\circ$ \\
 \multicolumn{1}{r}{ b} & $\theta^{(2)} > 80^\circ$ \\
 \multicolumn{1}{r}{ c} & $N^{(2)}_\mathrm{dir(b)} > 2$ \\ 
 2 & Filter 2:
  & 0.0232 & 0.0283 & 0.538 & 23.2 & 19.0 \\ 
 \multicolumn{1}{r}{ a} & $\theta^{(2)} > 90^\circ$ \\
 \multicolumn{1}{r}{ b} & $-0.43< S^{(2)}_\mathrm{mrl}< 0.3$ \\
 \multicolumn{1}{r}{ c} & $L^{(2)}_\mathrm{dir(c)} > 75$~m \\ 
 \multicolumn{1}{r}{ d} & $\mathcal{L}^{(2)}/\mathcal{L}^{(1)}<4\cdot10^{-6}$ \\
 \multicolumn{1}{r}{ e} & $\theta^{(5)} > 90^\circ$ \\ 
 3 & $\cos\theta^{(5)} < -0.1$ 
  & 0.867 & 0.861 & 0.925 & 1.07 & 1.07 \\ 
 4 & $\mathcal{P}^{(5)}_\mathrm{up}/\mathcal{P}^{(5)}_\mathrm{down} > 9.2$ 
  & 0.212 & 0.178 & 0.817 & 3.85 & 4.59 \\ 
 5 & $\mathcal{L}^{(5)}/\mathcal{L}^{(4)} < 1.02$
  & 0.370 & 0.334 & 0.947 & 2.56 & 2.84 \\ 
 6 & $-0.21< S_\mathrm{Phit}^{(3)} < 0.33$ 
  & 0.458 & 0.408 & 0.927 & 2.02 & 2.27 \\ 
 7 & $L^{(5)}_\mathrm{dir(c)} > 100$~m
  & 0.638 & 0.657 & 0.932 & 1.46 & 1.42 \\ 
 8 & $N^{(5)}_\mathrm{dir(c)}-N^{(4)}_\mathrm{dir(c)} > 3$ 
  & 0.737 & 0.710 & 0.912 & 1.24 & 1.28 \\ 
 9 & $-0.25<S^{(5)}_\mathrm{mrl}< 0.26$ 
  & 0.711 & 0.671 & 0.912 & 1.28 & 1.36 \\ 
 10& $L^{(5)}_\mathrm{dir(b)} > 40$~m 
  & 0.744 & 0.698 & 0.955 & 1.28 & 1.37 \\ 
 11& $\mathcal{P}^{(5)}/\mathcal{P}^{(4)} > 9.2$  
  & 0.581 & 0.562 & 0.921 & 1.59 & 1.64 \\ 
 12& $\mathcal{L}^{(3)} < 4.9$ 
  & 0.515 & 0.466 & 0.906 & 1.76 & 1.94 \\ 
 13& $N^{(5)}_\mathrm{dir(c)} > 9$
  & 0.659 & 0.616 & 0.963 & 1.46 & 1.56 \\ \tableline 
\end{tabular} 

\tablecomments{Describes selection criteria applied to reconstruction variables in the data reduction procedure.
For additional information on the filters and selection variables,
consult dissertation of S.~\citet{Young00}.
The numerical identification (superscript in parenthesis) refers to the reconstruction algorithm of the event:
(1) line fit used as first guess for likelihood fit,
(2) maximum likelihood method for muon track,
(3) hit probability reconstruction based on radial distribution of OMs which detect photons,
(4) maximum likelihood method assuming cascade event, and
(5) iterative application of maximum likelihood for muon track.
Direct hits are photons that arrive within (b) [$-15$,$+25$] ns, (c) [$-15$,$+75$] ns
of the unscattered time of flight between track and optical module.
The reduced likelihood parameter $\mathcal{L}$ is $-\log(\mathcal{P})$ divided by the number of
degrees of freedom, where $\mathcal{P}$ is the maximized probability.
Passing efficiencies ($\epsilon$) relative to the prior stage are shown at each stage for experimental data, simulated background, and simulated signal.  Rejection factors ($R$) for experimental data and simulated background are shown
for each stage in the two columns furthest to the right.\\
}
\end{table*}

\clearpage
\newpage
\begin{table*}
\caption{Muon and Neutrino Flux Limits on Selected Point Sources\label{table:limit}}

\normalsize
\begin{tabular}{ccccccc}
\tableline
\tableline
Source & Model & $N_{0}$ & $N_\mathrm{bg}$ & $\Delta E$ & $\Phi_{\nu}^\mathrm{limit}$ & $\Phi_{\mu}^\mathrm{limit}$ \\ 
&  &  & &[TeV]& [$10^{-8}$cm$^{-2}$s$^{-1}$] &  [$10^{-15}$cm$^{-2}$s$^{-1}$] \\ \tableline
Mkn 501 & 1 &  7 & 3.5& 0.3-20 & 86.0 & 38.9 \\
Mkn 501 & 2 &  7 & 3.5& 1-1000 & 9.5 & 14.6 \\
Mkn 421 & 3 &  4 & 3.7& 1-1000 & 11.2 & 9.7 \\
NGC4151 & 3 &  5 & 3.6& 1-1000& 12.9 & 10.9 \\
NGC4151 & 4 &  5 & 3.6& 60-2500 & 0.0042 & 5.6 \\
1ES2344 & 5 &  5 & 2.9& 1-400 & 12.5 & 10.3 \\ 
3C66A & 5 &  3 & 3.5& 0.8-250 & 7.2 & 6.6 \\
1ES1959+650 & 5 &  4 & 1.7& 0.8-250 & 13.2 & 9.7 \\
Crab Nebula & 5 &  2 & 5.6& 1-1000 & 4.2 & 5.0 \\
Cassiopeia A & 5 &  3 & 2.2& 1.8-1000 & 9.8 & 7.6 \\
Cygnus X-3 & 5 &  2 & 3.4& 1-1000 & 4.9 & 4.6 \\
Geminga & 5 &  4 & 7.1& 1.8-1000 & 6.8 & 9.1 \\ \tableline
\end{tabular}
\tablecomments{Muon and neutrino flux limits on selected sources for $E_{\nu}> 10$~GeV.
$N_{0}$ is the number of observed events in the search bin and $N_\mathrm{bg}$ is the expected background.
The energy interval, $\Delta E$, contains 90\% of the neutrino events, and the flux limits are corrected for systematic uncertainty (see Sec.~\ref{systematics}). Representative survey of models
(second column):
\textbf{1} - neutrino spectrum identical to measured photon spectrum \citep{HEGRA1997};
\textbf{2} - $d\Phi_{\nu}/dE \propto E_{\nu}^{-1.92}$;
\textbf{3} - \citet{SP92};
\textbf{4} - \citet{Stecker92};
\textbf{5} - $d\Phi_{\nu}/dE \propto E_{\nu}^{-2}$. \\
}    
\end{table*}

\clearpage
\newpage
\begin{table*}
\caption{Systematic Uncertainty in AMANDA-B10 Effective Area\label{tb:systematics}}  
\normalsize
\begin{tabular}{lc}
\tableline
\tableline
\rule[-2mm]{0mm}{7mm}{Source of systematic uncertainty} &
Error in $\overline{A}_\mathrm{eff}^{\mu}$ [$\pm$\%] \\ \tableline
Angular dependence of OM sensitivity             & 25    \\ 
Absolute OM sensitivity                          & 15    \\
Muon propagation                                 & 10    \\
Calibration (timing and geometry)                & 10    \\
Hardware simplifications in detector simulation  & $<10\;\;\;\;$ \\ 
Optical properties of bulk ice                   & 15    \\
\tableline
\end{tabular}
\end{table*}

\clearpage
\newpage
\begin{table*}
\caption{Muon Flux Limits\label{tb:allskyflux}} 

\scriptsize
\begin{tabular}{ccccccccc}  
\tableline
\tableline
Dec. & RA & $\Phi_{\mu}^\mathrm{limit}$ & Dec. & RA & $\Phi_{\mu}^\mathrm{limit}$ & Dec. & RA & $\Phi_{\mu}^\mathrm{limit}$ \\
\rm{[deg.]} & [hours]& [$10^{-15}$cm$^{-2}$s$^{-1}$] &[deg.]& [hours]& [$10^{-15}$cm$^{-2}$s$^{-1}$] & [deg.]& [hours]& [$10^{-15}$cm$^{-2}$s$^{-1}$] \\  \tableline

85	&4.0	&2.5	&  39 & 7.8 & 6.1 & 17 & 7.9 & 9.0\\
85	&12.0	&6.5	&  39 & 8.9 & 2,5 & 17 & 8.7 & 9.0\\
85	&20.0	&10.5	& 39 & 9.9 & 15.5 & 17 & 9.5 & 19.9\\
73	&1.3	&6.9	& 39 & 11.0 & 13.5 & 17 & 10.3 & 14.9\\
73	&4.0	&4.2	& 39 & 12.0 & 13.5 &17 & 11.2 & 42.6\\
73	&6.7	&2.1	& 39 & 13.0 & 8.0 & 17 & 12.0 & 11.9\\
73	&9.3	&6.9	& 39 & 14.1 & 8.0 & 17 &  12.8 & 34.1\\
73	&12.0	&13.7	& 39 & 15.1 & 8.0 & 17 & 13.7 & 24.6\\
73	&14.7	&6.9	 & 39 & 16.2 & 10.6 & 17 & 14.5 & 42.6\\
73	&17.3	&9.6	 & 39 & 17.2 & 15.5 & 17 & 15.3 & 8.9\\
73	&20.0	&4.2	& 39 & 18.3 & 3.9 & 17 & 16.1 & 11.9\\ 
73	&22.7	&4.2	& 39 & 19.3 & 6.1 & 17 & 17.0 & 14.9\\
62	&0.9	&7.9	& 39 & 20.3 & 10.6 & 17 & 17.8 & 8.9\\
62	&2.6	&5.5 &39 &  21.4 & 3.9 & 17 & 18.6 & 42.6\\
62	&4.3	&12.1	 & 39 & 22.4 & 3.9 & 17 & 19.4 & 19.9\\
62	&6.0	&3.4	 &39 & 23.5 & 6.1 & 17 & 20.3 & 19.9\\
62	&7.7	&7.9	 & 28 & 0.4 & 20.5 & 17 & 21.1 & 48.0\\
62	&9.4	&5.5	 & 28 & 1.3 & 20.5 & 17 & 21.9 & 14.9\\
62	&11.1	&7.9	 & 28 & 2.2 & 7.1 & 17 & 22.8 & 34.1\\
62	&12.9	&1.9	 & 28 & 3.1 & 13.9 & 17 & 23.6 & 29.1\\
62	&14.6	&5.5	 & 28 & 4.0 & 24.0 & 6 & 0.4 & 33.6\\
62	&16.3	&3.5	 & 28 & 4.9 & 13.9 & 6 & 1.2 & 64.9\\
62	&18.0	&3.5	 & 28 & 5.8 & 7.1 & 6 & 2.0 & 127.2\\
62	&19.7	&7.9	 & 28 & 6.7 & 20.5 & 6 & 2.8 & 64.9\\
62	&21.4	&7.9	 & 28 & 7.6 & 13.9 & 6 & 3.6 & 47.8\\
62	&23.1	&9.9	 & 28 & 8.4 & 20.5 & 6 & 4.4 & 33.6\\
51	&0.6	&7.7	 & 28 & 9.3 & 7.1 & 6 & 5.2 & 96.1\\
51	&1.9	&10.0	 & 28 & 10.2 & 24.0 &6 & 6.0 & 64.9\\
51	&3.2	&2.2	 & 28 & 11.1 & 20.5 & 6 & 6.8 & 78.0\\
51	&4.4	&7.7	& 28 & 12.0 & 24.0 & 6 & 7.6 & 78.0\\
51	&5.7	&5.9	& 28 & 12.9 & 5.4 & 6 & 8.4 & 47.8\\
51	&6.9	&10.0	& 28 & 13.8 & 13.9 & 6 & 9.2 & 64.9\\
51	&8.2	&2.2	& 28 & 14.7 & 5.4 & 6 & 10.0 & 64.9\\
51	&9.5	&10.0	& 28 & 15.6 & 20.5 & 6 & 10.8 & 127.2\\
51	&10.7	&5.9 & 28 & 16.4 & 16.7 & 6 & 11.6 & 47.9\\
51	&12.0	&1.6	 & 28 & 17.3 & 5.4 & 6 & 12.4 & 33.6\\
51	&13.3	&10.0	 & 28 & 18.2 & 16.7 & 6 & 13.2 & 143.5\\
51	&14.5	&5.9	 & 28 & 19.1 & 7.1 & 6 & 14.0 & 24.9\\
51	&15.8	&10.0	 & 28 & 20.0 & 10.2 & 6 & 14.8 & 47.8\\
51	&17.1	&7.7	 & 28 & 20.9 & 7.1 &6 & 15.6 & 33.6\\
51	&18.3	&14.1	 & 28 & 21.8 & 13.9 & 6 & 16.4 & 64.9\\
51	&19.6	&3.9	 & 28 & 22.7 & 10.2 & 6 & 17.2 & 47.8\\
51	&20.8	&5.9	 & 28 & 23.6 & 7.1 & 6 & 18.0 & 47.8\\
51	&22.1	&12.4	 & 17 & 0.4 & 48.0 & 6 & 18.8 & 112.4\\
51	&23.4	&7.7	 & 17 & 1.2 & 34.1 & 6 & 19.6 & 96.1\\
39	&0.5	&10.6	 & 17 & 2.1 & 14.9 & 6 & 20.4 & 33.6\\
39	&1.6	&15.5	 & 17 & 2.9 & 51.9 & 6 & 21.2 & 96.1\\
39	&2.6	&8.0	& 17 & 3.7 & 19.9& 6 & 22.0 & 24.9\\
39	&3.7	&13.5	& 17 & 4.6 & 24.6 & 6 & 22.8 & 64.9\\
39	&4.7	&3.9	& 17 & 5.4 & 14.9 & 6 & 23.6 & 78.0\\
39	&5.7	&15.5	& 17 & 6.2 & 4.5\\
39	&6.8	&6.1	& 17 & 7.0 & 9.0\\
\tableline
\end{tabular} 
\tablecomments{Neutrino-induced muon flux upper limits for source spectra proportional to $E^{-2}$.  The impact of systematic uncertainty is included. Angular coordinates refer to the center of the search bin.\\}
\end{table*}

\end{document}